\documentclass[
aps,
prc,
showpacs,
superscriptaddress,
groupedaddress]{revtex4-2}
\usepackage{graphicx}  
\usepackage{dcolumn}   
\usepackage{bm}        
\usepackage{amssymb}   
\usepackage{epstopdf}
\usepackage{amsmath}
\usepackage{nicefrac}
\usepackage{tikz}
\usepackage{enumitem}
\usepackage{comment}
\usepackage{array,multirow}
\usepackage[normalem]{ulem}
\usepackage{setspace}
\usepackage{hyperref}
\usepackage{soul}
\usepackage{lineno}
\usepackage[]{xcolor}
\newcommand {\nc} {\newcommand}
\nc {\IR} [1]{\textcolor{red}{#1}}
\nc {\IB} [1]{\textcolor{blue}{#1}}
\nc {\IP} [1]{\textcolor{magenta}{#1}}
\nc {\IM} [1]{\textcolor{Bittersweet}{#1}}
\nc {\IE} [1]{\textcolor{Plum}{#1}}
\nc {\IFB} [1]{\textcolor{ForestGreen}{#1}}

\begin{document}

\title{
Nonlocal
  \textit{in-medium} effective interaction for nucleon 
          scattering off isospin-asymmetric targets
 }
\author{J.~I. Fuentealba-Bustamante}
\affiliation{Department of Physics-FCFM, University of Chile, 
Av. Blanco Encalada 2008, RM 8370449 Santiago, Chile}
\affiliation{Department of Physics and Astronomy, Ohio University, 
Athens, Ohio 45701, USA}
\author{H.~F.~Arellano}
\affiliation{Department of Physics-FCFM, University of Chile, 
Av. Blanco Encalada 2008, RM 8370449 Santiago, Chile}

\date{\today}

\begin{abstract}
We have investigated the role of isospin asymmetry of 
the \textit{NN} effective interaction
in the context of \textit{NA} elastic scattering.
To this purpose we represent the \textit{in-medium} $g$ matrix 
as an admixture of isospin-symmetric nuclear matter 
and pure neutron matter solutions of Brueckner-Hartree-Fock 
equations for infinite nuclear matter,
denoted as $g[\rho,\beta]$.
We use the Argonne $v_{18}$ bare potential to 
represent the \textit{NN} interaction in free space,
due to its ability to 
describe the \textit{NN} scattering amplitudes up to 350~MeV.
The density-dependent isospin-asymmetric $g$ matrices 
are then used 
to calculate optical model potentials for elastic nucleon 
scattering off closed-shell nuclei.
For this aim, we make use of the
Arellano-Bauge
$\delta g$-folding approach 
suited for an explicit treatment of 
nonlocal density matrices.
This  approach allows to  account for the local isospin-asymmetry 
and density dependence
of the \textit{in-medium NN} interaction. 
The resulting optical potentials are nonlocal since 
the entire nonlocal structure of the $g$ matrix is retained.
We observe that including the isospin asymmetry in 
the $g$ matrix allows for a reasonable description 
of differential cross-sections at nucleon beam energies 
between 40 and 200~MeV.
%
In the case of proton scattering at energies below $\sim$65~MeV,
at momentum transfers $q$ below $\sim$1~fm$^{-1}$,
the inclusion of neutronic-matter $g$ matrices yields
better agreement with the data as compared to the case when 
symmetric nuclear matter $g$ matrices are used.
%
These results provide evidence that the isospin-asymmetry
in the \textit{NN} effective interaction
yield non-negligible effects in
nucleon scattering off isospin-asymmetric targets at
beam energies below 65~MeV.
\end{abstract}

\maketitle

\section{Introduction}

Current efforts in nuclear research are being directed toward 
the study of nuclear systems with significant isospin asymmetry, 
resulting in renewed interest in the study of nuclei at the limits 
of stability. Along this line, radioactive ion beam accelerators 
play a central role in delivering high-intensity beams of rare 
isotopes 
to collide against selected targets. 
Facilities such as 
FRIB at MSU in the USA, 
GSI in Germany,
EURISOL in Europe, 
ISOLDE at CERN, 
SPIRAL2 at GANIL in France, 
RIBF at RIKEN in Japan, 
and ISOL@MYRRHA in 
Belgium\cite{MSU,GSI,EURISOL,isolde,SPIRAL,RIKEN},
constitute examples of these developments. 
In the case of an exotic beam colliding with a hydrogen target, 
the process becomes equivalent to nucleon-nucleus 
(\textit{NA}) scattering. 
Indeed, inverse kinematics enables the
reduction of a, e.g., 65A-MeV ion scattering off a hydrogen target 
to a 65-MeV proton scattering off the nucleus of the ion. 
In this context, optical model potentials for \textit{NA} 
scattering constitutes a valuable tool for extracting 
information from scattering experiments involving rare isotope beams.

Considering that rare isotopes are exotic, featuring significant 
isospin-asymmetry, the detailed treatment of 
this effect must be guided by theory. Along
this line, microscopic optical potentials 
are still  a primary 
tool  to explore untested domains, avoiding 
phenomenological extrapolations. 
In this vein, significant developments have taken place 
such as \textit{ab-initio} calculations of optical potentials 
for low energy neutron 
scattering~\cite{Rotureau:2016jpf,Idini2018,Burrows:2023ygq} 
and \textit{ab-initio} folding potentials based on the Watson 
expansion~\cite{Burrows2019,Baker:2024wtn}. 
See Refs.~\cite{Hebborn2023,Dickhoff2019} for up-to-date 
reviews on the subject.
In this work we focus explicitly on accounting for the isospin asymmetry 
of the nuclear medium in the representation of 
the  \textit{in-medium} nucleon-nucleon (\textit{NN}) 
effective interaction.

The work we present here is an extension of the parameter-free
optical model introduced by Arellano and Bauge 
in 2007 \cite{Arellano2007a,Aguayo2008}, where the optical 
potential keeps track of a site-dependent effective interaction. 
In the model, Brueckner-Hartree-Fock (BHF) $g$-matrices have been used 
to represent the \textit{in-medium} \textit{NN} interaction,
although limited to the use of isospin-symmetric nuclear matter.

In the calculation of the optical potential for nucleon 
scattering from a given nucleus, the isospin asymmetry gets 
manifested by two elements. One of them comes from the 
one-body density matrix for the target protons and neutrons,
whereas the other stems
from the local isospin asymmetry at the site 
where the \textit{NN} interaction takes place. 
Current microscopic schemes 
for calculating microscopic optical potentials, make use 
of isospin-symmetric $g$ matrices, thus neglecting 
the asymmetry of the medium. 
We aim to go beyond this approach by accounting explicitly for the 
isospin asymmetry in the \textit{NN} interaction.  
By incorporating explicitly $g$ matrix 
solutions for pure neutron (proton) matter, 
the $g$ matrix becomes a functional of the isoscalar 
density $\rho$ and isospin asymmetry $\beta$, 
to be expressed as $g[\rho,\beta]$. 

Investigations on the density and asymmetry dependence of $g$ 
matrices within BHF approximation have been reported in 
Refs. \cite{Bombaci1991,Song1992,Zuo1999}.
For this work, we refer to the article
by Bombaci and Lombardo, where they report BHF calculations of 
considering various asymmetries $\beta\!=\!(N-Z)/A$.
The interesting finding they report is that
the depth of the single-particle (s.p.) potentials for protons and
neutrons vary nearly linearly as functions of $\beta$, with
$0\!\leq\!\beta\!\leq\!1$.
Since the s.p. self-consistent
potentials correspond to the on-shell mass operator,
which in turn scales with the $g$ matrix, 
we can then assume that the $g$ matrix varies 
--to lowest order-- linearly as a function of $\beta$. 
We apply this linear behavior in the construction of the $g$ matrix,
and interpolate the solutions for symmetric nuclear matter
($\beta\!=\!0$) and pure neutron matter ($\beta\!=\!1$) linearly.
This construction enables us to assess the role of the isospin
asymmetry in the microscopic calculation of optical potentials
for isospin-asymmetric targets such as $^{48}$Ca, $^{90}$Zr and
$^{208}$Pb. We will consider proton and neutron probes at energies
up to 200~MeV. The bare interaction to be used is
Argonne $v_{18}$ (AV18)~\cite{Wiringa1995}.

This article is organized as follows. 
In Section 2, we present the theoretical framework, where we review 
general aspects of the $\delta g$ folding. 
We also introduce a model incorporating the isospin
asymmetry in the \textit{NN} effective interaction. 
In Section 3, we present the main considerations for the implementation 
of the $\delta g$ folding. 
We also briefly outline the \textit{in-medium} $g$ matrix in BHF.
In Section 4 we present results for proton and neutron elastic
scattering from the isospin-asymmetric targets
$^{208}$Pb and $^{48}$Ca. 
Section 5 summarizes the work and draws its main conclusions.

\section{Framework}
The optical potential for \textit{NA} scattering can be 
formulated in multiple ways. Most of them  can be reduced 
to the expectation value of a generalized two-body effective 
interaction with the nucleus in its
ground 
state~\cite{Watson1953,Feshbach1962,Kerman1959,Watson1965,Villars1967}. 
With this observation, the optical potential in momentum 
space for the scattering of a nucleon of energy $E$ from 
a  nucleus can be expressed as
\begin{equation}
     U(\boldsymbol{k'},\boldsymbol{k};E) = 
     \int d\boldsymbol{p'} \, d\boldsymbol{p} \:
     \langle \,\boldsymbol{k'} \, \boldsymbol{p'} \,| \,
\hat{T}(E) \, 
  | \, \boldsymbol{k} \, \boldsymbol{p} \,\rangle_{\mathcal{A}}  \,  
  \hat{\rho}(\boldsymbol{p'},\boldsymbol{p}), \label{ompot}
\end{equation}
where the subscript $\mathcal{A}$ denotes
antisymmetrization and $\hat{\rho}(\boldsymbol{p'},\boldsymbol{p})$ 
is the nonlocal one-body density matrix of the target. 
In this way, the information about the multi-nuclear 
nature of the system is contained in the $\hat{T}$ matrix. 
In the above, $\boldsymbol{k}$ and $\boldsymbol{p}$ 
represent the momenta of the projectile and struck nucleon before 
the interaction, respectively. 
Analogously, $\boldsymbol{k'}$ and $\boldsymbol{p'}$ refer 
to the post-interaction momenta. 

In quite general terms, it is shown in Ref. \cite{Arellano2007a} 
that any two-body 
operator $\hat{T}$ can be expressed in the following functional form
\begin{equation}
     \langle \,\boldsymbol{k'} \, \boldsymbol{p'} \,| \, \hat{T} \, 
     | \, \boldsymbol{k} \, \boldsymbol{p} \,\rangle = 
     \int \frac{d\boldsymbol{z}}{(2\pi)^3}
     e^{i\boldsymbol{z} \cdot (\boldsymbol{Q}-\boldsymbol{q})}
     g_{\boldsymbol{z}}(\boldsymbol{K} + 
     \boldsymbol{P};\boldsymbol{b'},\boldsymbol{b}),
     \label{Tk1}
\end{equation}
where $g_z$ is a reduced matrix which depends on the $z$ coordinate.
The above result comes after transforming 
$\langle {\bm k'}{\bm p'}| \hat{T} | {\bm k}\,{\bm p}\rangle$
to coordinate space,
obtaining 
$\langle {\bm r'}{\bm s'}| \hat{T} | {\bm r}\,{\bm s}\rangle$,
and then transforming it back to the momentum representation.
In the process, ${\bm z}$ is identified as the average coordinate
of both prior and post coordinates, namely
\begin{equation}
  \label{zmean}
  {\bm z}=\frac{1}{4}({\bm r}+{\bm s}+{\bm r'}+{\bm s'})\;.
\end{equation}
Additionally,
\begin{equation}
  \label{relative}
{\bm b'}= \textstyle{\frac{1}{2}}({\bm k'}-{\bm p'})\,,
\qquad
{\bm b}= \textstyle{\frac{1}{2}}({\bm k}-{\bm p})\,,
\end{equation}
represent the pre- and post-collision relative momenta.
Furthermore,  
\begin{equation}
  \label{Kq}
{\bm K} = \textstyle{\frac{1}{2}} ({\bm k} + {\bm k'})\,,
\qquad
{\bm q}={\bm k} - {\bm k'}\,,
\end{equation}
represent the 
projectile average momentum
and the momentum transfer.
Likewise,
\begin{equation}
  \label{PQ}
{\bm P} = \textstyle{\frac{1}{2}} ({\bm p} + {\bm p'}) \,,
\qquad
{\bm Q} = {\bm p'} - {\bm p}\,,
\end{equation}
account 
for the struck-nucleon average momentum and the recoil momentum.

Guided by folding models using density-dependent \textit{NN} 
effective interactions~\cite{Brieva1977b,Geramb1983,Amos2000},  
the authors in Ref. \cite{Arellano1995} assume $g_z$ 
as the fully off shell \textit{in-medium} $g$ matrix for 
infinite nuclear matter. The density at which the $g$ matrix
is evaluated correspond to that at the coordinate $z$, namely $\rho(z)$.
After replacing Eq. \eqref{Tk1} for $\hat{T}$ 
in Eq. \eqref{ompot} for $U$, we obtain 
\begin{align}   
  \nonumber
  U({\bm k'},{\bm k};E) =&
    \frac{1}{(2 \pi)^3}\int d{\bm Q}\; d{\bm P}\; 
    d{\bm z}\; 
    e^{i{\bm z} \cdot ({\bm Q} - {\bm q})}
    \times \\
   &g_z({\bm K} \!+\! {\bm P}; 
    {\bm b'},{\bm b})
   \:\hat{\rho}
   ({\bm Q};{\bm P})
\end{align}
As observed, the optical potential involves multidimensional 
integrals, namely three-dimensional integrals over
 $\boldsymbol{z}$, $\boldsymbol{Q}$ and $\boldsymbol{P}$.
In Ref. \cite{Arellano2011b}, it was shown that a reasonable 
approximation to the above expression for $U$ is 
setting ${\bm Q} \!=\!{\bm q}$ in the $g$ matrix, 
so that
the relative momenta become
\begin{equation}
    {\bm b'} = \frac12({\bm K} - {\bm P} - {\bm q});\qquad
    {\bm b} = \frac12({\bm K} - {\bm P} + {\bm q}).
\end{equation}
Additionally, the authors of Ref. \cite{Arellano2007a} 
demonstrate that the optical potential can be expressed
as the sum of two terms in the form
\begin{equation}
  U({\bm k'},{\bm k};E)
  = U_0({\bm k'},{\bm k};E) + U_1({\bm k'},{\bm k};E), \label{eq. u0+u1}
\end{equation}
where
\begin{subequations}
\begin{align}
  U_0({\bm k'},{\bm k};E) =&
    \int d{\bm P}\,
    \langle\, {\bm b'}\,|\,
    t\, |\, {\bm b}\,\rangle_{\mathcal{A}} \;
    \hat{\rho}({\bm q};{\bm P})\;
    ;\label{U0} \\
    \nonumber
    U_1(\boldsymbol{k'},\boldsymbol{k};E) =&
    \frac{-1}{6\pi^2} \int_0^{\infty}\!\! z^3 dz     
    \int d{\bm P} 
    \int d\boldsymbol{Q}\,\times
    \\
    & 
    \left [
      \frac{3\,j_1 (|{\bm Q} - {\bm q}|z)}{|{\bm Q}-{\bm q}|z}
      \right ]
      \left \langle\, {\bm b'}\, \right |
     \frac{\partial g_z}{\partial z}
     \left | \, {\bm b}\, \right \rangle_{\mathcal{A}} 
     \, \hat{\rho}(\boldsymbol{Q};\boldsymbol{P}). 
     \label{U1}
\end{align}
\end{subequations}
Here, $j_1$ denotes the spherical Bessel function of first order.
In calculations reported here, both $t$ and $g_z$ are evaluated at 
an average value of the pair momentum ${\bm K}\!+\!{\bm P}$.
Note that the intrinsic medium effects get manifested as the 
gradient of $g_z$ at the (mean) average coordinate $z$.
This feature of the model motivates to refer to it
as $\delta g$ folding. 

Note that if the reduced interaction $g_z$ is 
translationally invariant, as is the case for the free $t$ matrix, 
then $\partial g_z/ \partial z$=0, so that $U_1$ vanishes. 
The resulting potential reduces to $U_0$, corresponding 
to the medium-free full folding potentials reported in
Refs.~\cite{Arellano1989,Arellano1990,Elster1990,Crespo1990,Vorabbi2024}.

An interesting property which emerges from the above 
expression for $U$ is that if the density matrix is expressed 
in the Slater approximation (See appendix A), then $U$ gets reduced 
to~\cite{Arellano2007a}
\begin{equation}
  U_{\textrm{ABL}}({\bm k'},{\bm k};E) =
    4 \pi \int_0^{\infty} z^2\: dz\: j_0(qz) \:\rho(z) 
    \langle g_z^{(0)}({\bm k'},{\bm k}) \rangle. 
    \label{UABLth}
\end{equation}
In this expression
$\langle \!\ g_z({\bm k'},{\bm k}) \rangle$ 
represents the Fermi-motion integral 
\begin{equation}
    \langle  \ g_z({\bm k'},{\bm k}) \rangle =
    \int d\boldsymbol{P} \:n_z(P)\:
    g_z({\bm K}\!+\!{\bm P};{\bm b'},{\bm b})\:.
\end{equation}
Here, $n_z(P)$ is given by Eq.~\eqref{nzP},
setting bounds for the off-shell sampling of the
interaction $g_z$ at density $\rho(z)$.
From now on, we shall refer to this expression for the optical 
potential as ABL folding, in reference to its 
authors~\cite{Arellano1995}. 
This expression has become useful for 
us to check the subroutines coded in this work. 
The $g$ matrix is evaluated at a starting 
energy $\omega\!=\!E\!+\!\Bar{e}$, with $E$ being the kinetic 
energy in the \textit{NA} center of mass (c.m.), 
and $\Bar{e}$ an average nuclear-matter s.p. energy 
at the local density $\rho(z)$.

To incorporate the isospin asymmetry in the $g$ matrix, 
we proceed as follows. Let $\rho_p(z)$ and $\rho_n(z)$ be the 
point proton and neutron densities, respectively.
Then, we define the isoscalar density as 
\begin{equation}
    \rho(z) = \rho_p + \rho_n.
\end{equation}
Additionally, to characterize the isospin asymmetry at each 
radial site $z$ in the nucleus we define the local isospin 
asymmetry $\beta(z)$ as 
\begin{equation}
    \beta(z) = \frac{\rho_n - \rho_p}{\rho_n + \rho_p}.
\end{equation}
Considering the density-dependent $g$ matrix 
a functional of the isoscalar density $\rho$ and 
isospin asymmetry $\beta$, then $g\!=\!g[\,\rho,\beta\,]$.
As justified in Sec. I, the dependence
of the $g$ matrix on the asymmetry is assumed to be linear.
Thus, if $g^{(0)}[\rho]$ denotes the $g$ matrix in symmetric nuclear
matter (SNM), and
$g^{(1)}[\rho]$ denotes the one for pure neutron matter (PNM),
then functional dependence of $g$ on $\rho$ and $\beta$
gets expressed as
\begin{equation}
    g[\rho,\beta] = g^{(0)}[\,\rho\,] \; (1-\beta) + g^{(1)}[\,\rho\,]
    \; \beta .\label{g_rb}
\end{equation}
Note that if the medium is nearly isospin symmetric, 
as in $^{40}$Ca, $\beta\!\approx\!0$, so that
$g[\rho,\beta]\! \approx\! g^{(0)}[\,\rho\,]$. 
In the other extreme, such as in neutron stars, 
then $\beta\! \approx\! 1$,
so that $g[\rho, \beta] \!\approx\! g^{(1)}[\rho]$.

To calculate $U_1$ from Eq.~\eqref{U1} 
we need to evaluate the gradient term $\partial g/\partial z$ 
at each coordinate $z$. Consistent with Eq.~\eqref{g_rb} we get 
\begin{equation}
     \frac{\partial g}{\partial z} =
     \left[  \frac{\partial g^{(0)}}{\partial z} \; (1-\beta) - g^{(0)} \;
     \frac{\partial \beta}{\partial z}\right] + 
     \left[\frac{\partial g^{(1)}}{\partial z} \; \beta + g^{(1)} \; 
     \frac{\partial \beta}{\partial z}\right],
     \label{dgzb}
\end{equation}
where
\begin{subequations}
\begin{align}
    \frac{\partial g^{(0)}}{\partial z} &= 
    \frac{\partial g^{(0)}[\rho]}{\partial \rho} \; 
    \frac{\partial \rho (z)}{\partial z}, \\
    \frac{\partial g^{(1)}}{\partial z} &=
    \frac{\partial g^{(1)}[\rho]}{\partial \rho} \; 
    \frac{\partial \rho (z)}{\partial z}.
\end{align}
\end{subequations}

\begin{center}
\begin{figure}
  \includegraphics[width=0.8\linewidth,angle=-90]
   {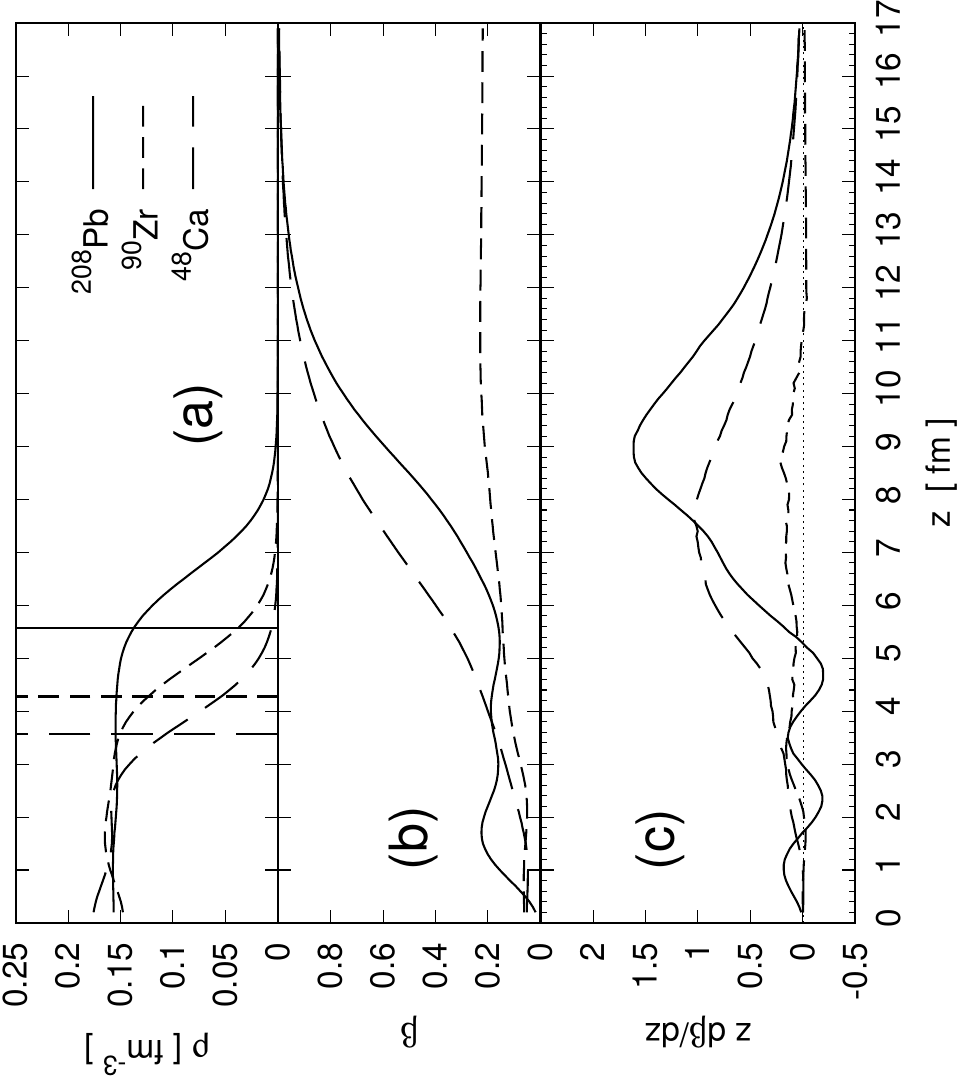}
\caption{
  \label{fig:RhoDBeta}
Isoscalar density $\rho$, isospin asymmetry $\beta$ and 
dimensionless gradient $z(\partial\beta/\partial z)$
as functions of the radial 
coordinate $z$ for $^{208}$Pb, $^{90}$Zr, and $^{48}$Ca.
The vertical lines in panel (a) denote the rms radius for each nucleus.}
\end{figure}
\end{center}

To illustrate the behavior of the isospin-asymmetry $\beta(z)$ 
for nuclei of interest, in Fig.~\ref{fig:RhoDBeta} we show 
the isoscalar density $\rho$ (a), the
isospin-asymmetry $\beta$ (b) and its dimensionless 
gradient $z(\partial\beta/\partial z)$ as functions of the radial
coordinate $z$. Solid, short-dashed, and long-dashed curves 
correspond to $^{208}$Pb, $^{90}$Zr and $^{48}$Ca, respectively. 
The vertical lines in panel (a) denote the root-mean-squared (rms)
radius for each nucleus. 

We observe that the isoscalar point densities $\rho(z)$ 
displayed in panel (a) are nearly constant up to about their 
respective rms radius. Beyond that point the density
decreases rapidly as the radial coordinate increases. 
It is also interesting to note that both $\beta$ 
and $z(\partial\beta /\partial z)$ for $^{90}$Zr are rather small
compared to the other two nuclei.
This feature points to that
the role of $g^{(1)}[\rho]$ in the 
optical potential for $^{90}$Zr as target would be marginal,
as inferred from Eq.~\eqref{dgzb} for $U_1$.

\section{Implementation}

\subsection{Target ground state}
For the actual calculation of the optical potential, we need to describe 
the target in its ground state. For this purpose, we use shell-model 
single-particle (s.p.) wavefunctions based 
on a computational code furnished by J. Negele \cite{Negele1970}. 
These correspond to self-consistent Hartree-Fock calculations
in the local density approximation.

\begin{center}
  \begin{figure}[h!]
    \includegraphics[width=0.9\linewidth,angle=-90]
    {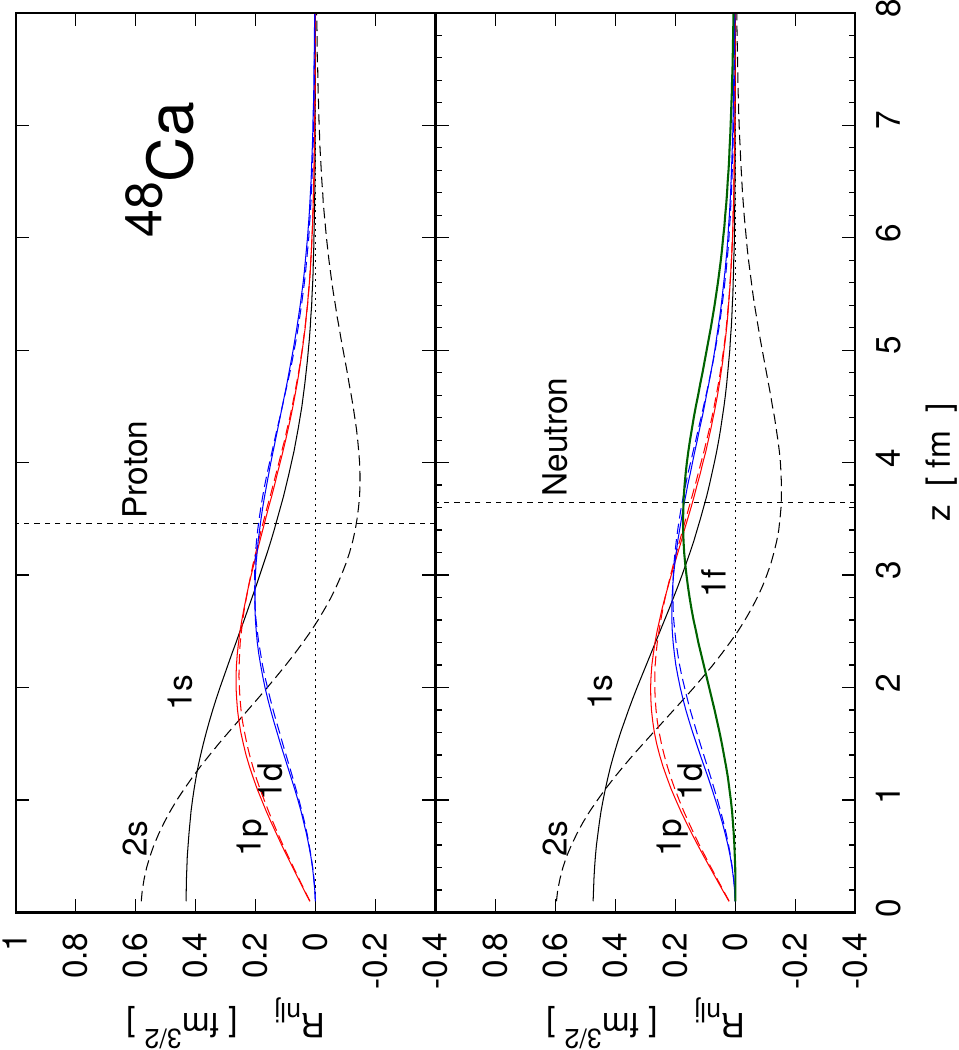}
\caption{ \label{fig:wfCa48}
    Single-particle wavefunctions $R_{nlj}$ for $^{48}$Ca as
  functions of the radial coordinate $z$. 
    Solid curves denote states with $j\!=\!|l\!-\!1/2|$, whereas
    dashed curves denote those with $j\!=\!l\!+\!1/2$.
}
  \end{figure}
\end{center}

In Fig. \ref{fig:wfCa48} we show the s.p. wavefunctions $R_{nlj}(z)$
as functions of the radial coordinate $z$ for $^{48}$Ca. 
Here $n$ denotes the principal quantum number, $l$ the orbital 
angular momentum and $j$ the total angular momentum. 
This is a spherical nucleus composed of twenty protons and 
twenty-eight neutrons. 
As a result, there are six and seven shells, respectively. 
Black, red, blue, and green curves denote $s$, $p$, $d$ 
and $f$ waves. 
Note that only the $s$ shell has two bound states, with 
the second one featuring a node around 2.5~fm (black dashed curves). 
The root-mean-squared radii for proton and neutron
densities are $3.46$ and $3.65$~fm, respectively. 

In terms of the s.p. wavefunctions of the nucleus,
the point density is given by 
\begin{equation}
    \rho(z) = \sum_{nlj}(2j+1)|R_{nlj}(z)|^2,
\end{equation}
with the summation over proton or neutron shells.
For the nonlocal density matrix we adopt the following construction
suited for closed-shell nuclei
\begin{equation}
    \rho({\bm r'},{\bm r}) = 
    \sum_{nlj}(2l+1) R_{nlj}(r') R_{nlj}(r) 
    P_{l}(\hat{\bm r}' \cdot \hat{\bm r}),
\end{equation}
with $P_l$ the Legendre polynomial of order $l$.
For our purposes, it is convenient to represent the 
density matrix in momentum space. Therefore, we perform a 
double Fourier transform resulting in
\begin{align}
  \Tilde{\rho}({\bm p'},{\bm p}) &= 
    \frac{1}{(2\pi)^3} \int d{\bm r}\:d{\bm r'}\:
    e^{i({\bm r'} \cdot {\bm p'} - {\bm r} \cdot {\bm p} )}
    \; \rho({\bm r'},{\bm r}) \nonumber \\
     &= 
    \sum_{nlj} (2j+1) \Tilde{R}_{nlj}(p')\, \Tilde{R}_{nlj}(p)\,
    P_l(\hat{\bm p}'\cdot \hat{\bm p}),
\end{align}
where
\begin{equation}
    \Tilde{R}_{nlj}(p) = \sqrt{\frac{2}{\pi}}\int_0^{\infty}
    r^2 dr\: j_l(pr)\:  R_{nlj}(r),
\end{equation}
with $j_l$ the spherical Bessel function of the first kind, of order $l$.
Let us now introduce $\boldsymbol{P} =
\textstyle{\frac{1}{2}}(\boldsymbol{p} + \boldsymbol{p'})$, 
and $\boldsymbol{Q} = \boldsymbol{p'} - \boldsymbol{p}$, 
so that 
$\boldsymbol{p'} = 
\boldsymbol{P} + \textstyle{\frac{1}{2}}\boldsymbol{Q}$, 
and $\boldsymbol{p} =
\boldsymbol{P} - \textstyle{\frac{1}{2}}\boldsymbol{Q}$. 
Then
\begin{align}
    \Tilde{\rho}(\boldsymbol{p'},\boldsymbol{p}) &= 
    \Tilde{\rho}(\boldsymbol{P} + \textstyle{\frac{1}{2}}\boldsymbol{Q},
    \boldsymbol{P} - \textstyle{\frac{1}{2}}\boldsymbol{Q}), \\
    &\equiv \Tilde{\rho}(\boldsymbol{Q};\boldsymbol{P}).
\end{align}
With this choice, the density matrix depends on the two vector
momenta
$\boldsymbol{P}$ and $\boldsymbol{Q}$, 
where $\boldsymbol{P}$ accounts for the commonly known Fermi motion, 
typically bound to about $1.5$~fm$^{-1}$. 
If this momentum is summed over,  we obtain the Fourier 
transform of the radial density $\rho(z)$, namely 
\begin{equation}
    \int \Tilde{\rho}(\boldsymbol{Q};\boldsymbol{P}) 
    d\boldsymbol{P} = \Tilde{\rho}(\bm Q),
\end{equation}
which at $Q\!=\!0$ is normalized to the number of nucleons.

In this work, we find it useful to express the density matrix as a 
Legendre expansion in terms of
$w=\hat{\boldsymbol{Q}}\cdot \hat{\boldsymbol{P}}$ . 
Thus
\begin{equation}
    \Tilde{\rho}(\boldsymbol{Q};\boldsymbol{P}) = 
    \sum_{n} \Tilde{\rho}_n(Q,P)P_n(w).
\end{equation}
In practice,
\begin{equation}
    \Tilde{\rho}_n(Q,P) = \left( n+\frac{1}{2}\right)  
    \;    \int_{-1}^{1} \Tilde{\rho}(\boldsymbol{Q};\boldsymbol{P}) 
    \, dw, \label{pmultipole}
\end{equation}
with $w=\hat{\boldsymbol{Q}}\cdot \hat{\boldsymbol{P}}$, 
corresponding to the cosine of the angle 
between $\boldsymbol{Q}$ and $\boldsymbol{P}$.

\begin{center}
\begin{figure}[ht]
  \includegraphics[width=0.9\linewidth]
  {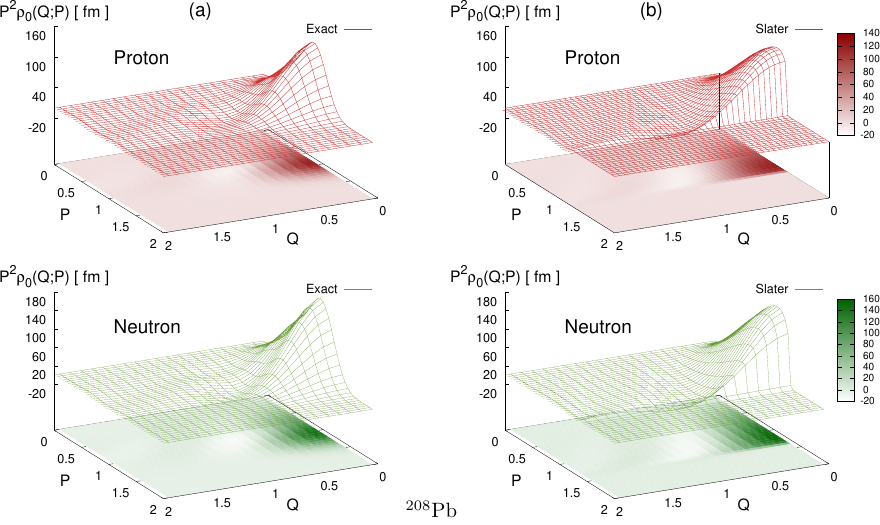 }
    \caption{
    \label{fig:rhoCa48}
      Surface plots in the $QP$ plane for proton 
   and neutron 
   densities in $^{208}$Pb. 
    Surfaces on the lhs
    correspond to the exact monopole density matrix, 
 whereas on the rhs
  correspond to the Slater approximation.}
\end{figure}
\end{center}

In Fig.~\ref{fig:rhoCa48} we present surface plots in the $QP$ plane of 
the density matrix for $^{208}$Pb. On the left-hand side (lhs) 
we plot the monopole term ($n=0$). On the right-hand side (rhs) 
we show results for the density matrix 
in the Slater approximation \cite{Arellano1990b}
(see Appendix A).
Red and green
surfaces denote proton and neutron densities, respectively. 
We observe slight differences between the monopole (lhs) and
Slater approximation (rhs) in their $P$ dependence,
with the Slater approximation featuring 
vanishing contributions above a given momentum $P_{max}$.  

\subsection{\textit{In-medium} $g$ matrix}

In order to represent the \textit{NN} effective 
interaction at each radial coordinate $z$ in the nucleus, 
we resort to the $g$ matrix in the BHF approximation. 
This concept is an extension of the local density approximation 
proposed by Brieva and Rook \cite{Brieva1977a}. 
In this work, we account for all non-localities of the interaction
by retaining the intrinsic off-shell structure of the $g$ matrix. 
Furthermore, at each coordinate $z$, we include the dependence of
the $g$ matrix on the local isoscalar density $\rho(z)$ and 
isospin asymmetry $\beta(z)$.

For infinite SNM or infinite PNM, the $g$ matrix in the BHF approximation
depends on the density $\rho$ of the medium characterized by the Fermi 
momentum $k_F$ and a starting energy $\omega$. 
In this limit, when two-body correlations in the ladder
approximation are taken into account, the $g$ matrix satisfies
\begin{equation}
    \hat{g}(\omega) = \hat{v} + \hat{v} \;
    \frac{\hat{Q}}{\omega + i\eta -\hat{h}_1 - \hat{h}_2}\hat{g}(\omega),
    \label{eq.bhf}
\end{equation}
with $\hat{v}$ being the bare interaction between nucleons. 
Here $\hat{h}_{1}$ and $\hat{h}_{2}$ denote the s.p. energy of nucleons $1$ and $2$.
The Pauli blocking operator $Q$ satisfies 
\begin{equation}
    \hat{Q}
    |\, \boldsymbol{p} \,\boldsymbol{k}\,\rangle =
    \Theta(p-k_F)\;\Theta(k-k_F)\,
    |\,\boldsymbol{p}\, \boldsymbol{k}\,\rangle. \label{NMS-01}
\end{equation}
Here $\Theta(x)$ represents the Heaviside step function. 
The solution to Eq. \eqref{NMS-01} enables the evaluation
of the mass operator
\begin{equation}
    M(k;E) = 
    \sum_{|\boldsymbol{p}|\leq k_F} 
    \langle \,\textstyle{\frac{1}{2}}(\boldsymbol{k} - \boldsymbol{p}) \, |g_{\boldsymbol{K}} (E + e_p )| \, \textstyle{\frac{1}{2}}(\boldsymbol{k} -\boldsymbol{p})\rangle_{\mathcal{A}}, 
\end{equation}
where $\boldsymbol{K}$ is the total momentum of the 
interacting pair, $\boldsymbol{K}=\boldsymbol{k}+\boldsymbol{p}$, and 
\begin{equation}
    e_p = \frac{p^2}{2m} + U(p),
\end{equation}
the s.p. energy defined in terms of an auxiliary field $U$.
The nucleon mass $m$ is taken as the average of proton and 
neutron masses. In the BHF approximation the s.p. potential 
is given by the real part of the on-shell mass operator, namely
\begin{equation}
    U(k) = \Re\textrm{e}\, M(k;e_k).
\end{equation}
This imposes a self-consistency requirement, a condition which 
is achieved iteratively. In the continuous choice, this condition 
is imposed to all momenta $k$ \cite{Baldo2000}. 
Further details on the way BHF solutions are obtained for
this work can be found in Ref. \cite{Arellano2015}.

In this study, we employ the AV18 \textit{NN} 
potential \cite{Wiringa1995} to represent the bare interaction. 
Solutions for the corresponding SNM and PNM 
have been reported in Refs. \cite{Arellano2015, Isaule2016}. 
In this case, the saturation point of SNM occurs
at $k_F=1.55$ fm$^{-1}$. For these solutions, we obtain a 
slope parameter $L=56.67$ MeV, a value consistent with those 
reported by Danielewicz in Ref. \cite{Danielewicz2009}. 
The slope parameter is calculated from
\begin{equation}
    L= 3 \rho_0
    \frac{\partial S(\rho)}{\partial \rho}\bigg\rvert_{\rho=\rho_0}.
\end{equation}
 with $\rho_0$ the saturation density. Here, $S(\rho)$ 
 denotes the symmetry energy.

\begin{figure}[ht]
\begin{center}
\begin{tikzpicture}
\node[inner sep=0pt] (sun) at (-4.5,0){\includegraphics[width=.32\textwidth,angle=-90]{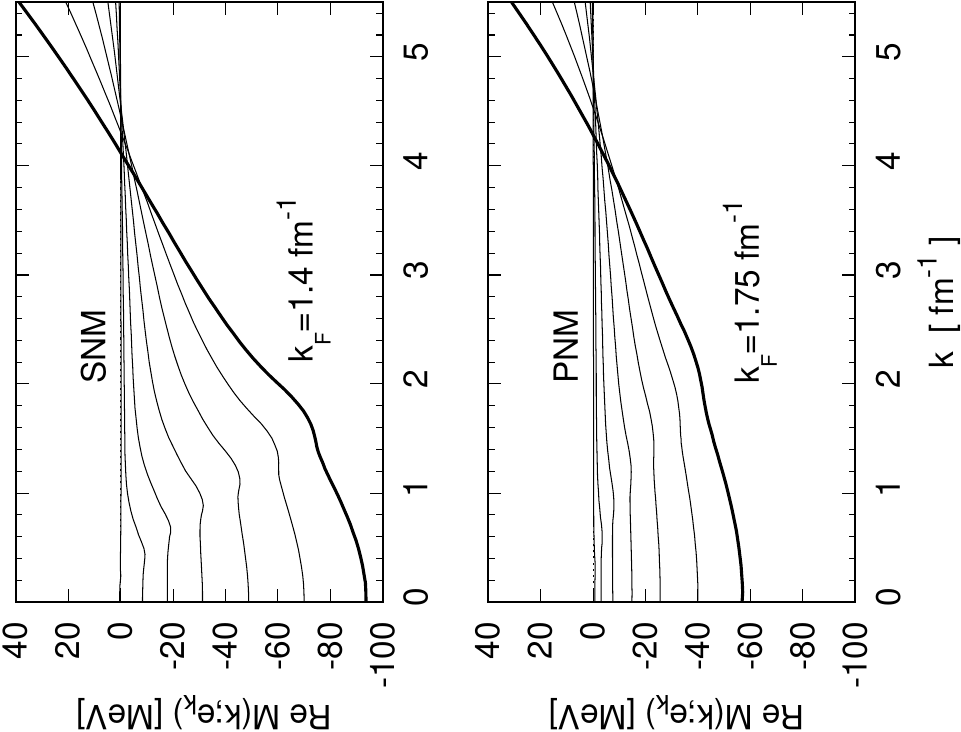}};
\node[inner sep=0pt] (earth) at (0,0){\includegraphics[width=.32\textwidth,angle=-90]{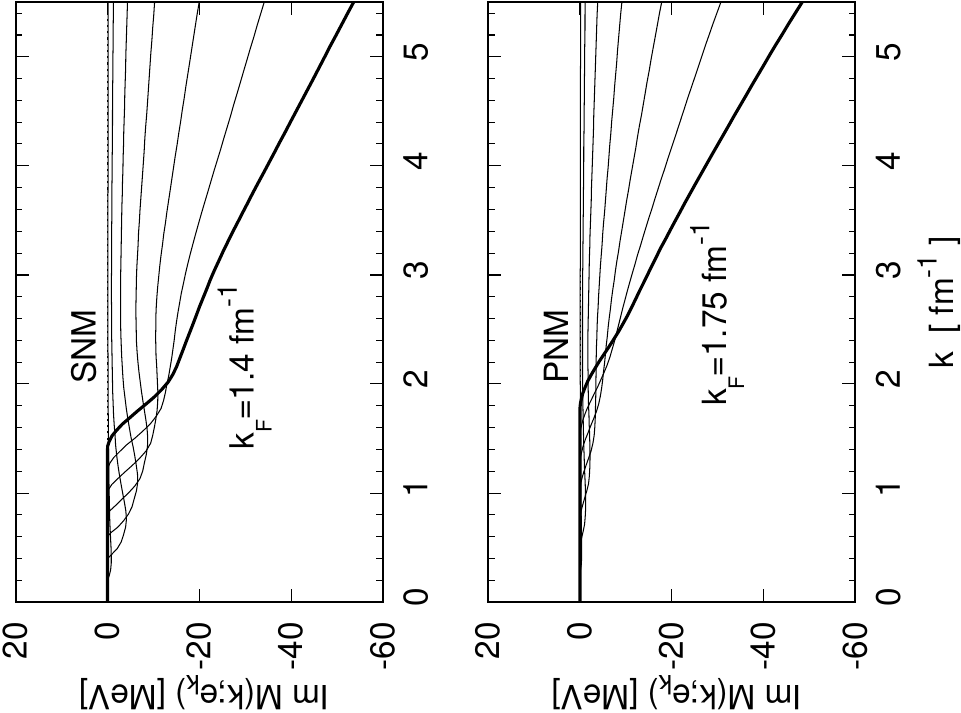}};
\end{tikzpicture}
\end{center} 
    \caption{
    On-shell mass operator 
    $M(k;e_k)$
    at different $k_F$ as functions of $k$.
    Curves for SNM (upper frames) denote results
    with $0.4\!\leq\!k_F\!\leq\!1.4$~fm$^{-1}$ in steps of 0.2~fm$^{-1}$,
  while those for PNM (lower frames) correspond to 
   $0.5\!\leq\!k_F\!\leq\!1.75$~fm$^{-1}$ in steps of 0.25~fm$^{-1}$.}
    \label{fig:ReU_ImU}
\end{figure}

An appealing advantage of using the AV18 potential is that 
it describes \textit{NN} observables up to energies of about $350$~MeV. 
In Fig. \ref{fig:ReU_ImU} we plot the real (lhs) and imaginary part )rhs)
of the on-shell mass operator based on AV18.
We observe that the depth of the real part of the mass operator 
increases with the density. 
Additionally in both cases $U(k)$ crosses the axis around 
$k\!\approx\!4$~fm$^{-1}$.
The imaginary part, on the other hand, vanishes for $k\!\leq\!k_F$, 
as expected.
The thick solid curves correspond to s.p. solutions near 
 the nuclear saturation density $\rho_0\!=\!0.16\textrm{fm}^{-3}$. 
We note that $U(k\!=\!0)$ for PNM is weaker by about 40\% 
relative to SNM.
These s.p. potentials are the ones used in Eq. \eqref{eq.bhf} to obtain
the off-shell elements
$\langle {\bm\kappa'} | g(\omega;k_F)| {\bm\kappa} \rangle$ 
needed to evaluate $U({\bm k'},{\bm k})$ 
in Eqs.~\eqref{U0} and \eqref{U1}. 

\subsection{Spin-isospin considerations}
Here, we outline some general considerations made to 
incorporate the
spin-isospin degrees of freedom in the \textit{NN} effective
interaction, which is generally denoted by $t$.
In general terms, this operator can be expressed as
\begin{equation}
	\label{t1}
t = t_0 + 
    t_\tau   ({\bm\tau}_1\cdot{\bm\tau}_2) +
    t_\sigma ({\bm\sigma}_1\cdot{\bm\sigma}_2) + 
t_{\sigma\tau}({\bm\sigma}_1\cdot{\bm\sigma}_2)({\bm\tau}_1\cdot{\bm\tau}_2)\;,
\end{equation}
where ${\bm\sigma}$ and ${\bm\tau}$ denote Pauli matrices in spin
and isospin spaces, with the subscripts $1,2$ referring to each nucleon.
Here $t_0$ and $t_\tau$ denote the isoscalar and isovector components of 
the force.
The total spin and isospin operators read
\begin{subequations}
\begin{align}
  \mathbf{S} &= \tfrac{1}{2}{\bm\sigma}_1  +
              \tfrac{1}{2}{\bm\sigma}_2, \\
  \mathbf{T} &= \tfrac{1}{2}{\bm\tau}_1  +
              \tfrac{1}{2}{\bm\tau}_2, 
\end{align}
\end{subequations}
respectively.
Thus, for the \textit{NN} system the total spin $S$ can be zero or one.
The same for the isospin ($T=0,1$). 
In this basis each of the four components in Eq.~\eqref{t1} is given by
\begin{subequations}
\begin{align}
  t_0  &= \textstyle{\frac{1}{16}}
  ( T^{00} + 3 T^{01} + 3 T^{10} + 9 T^{11} ), \\
  t_{\tau} &= \textstyle{\frac{1}{16}}
  (-T^{00} +   T^{01} - 3 T^{10} + 3 T^{11} ), \\
  t_{\sigma} &= \textstyle{\frac{1}{16}}
  (-T^{00} - 3 T^{01} +   T^{10} + 3 T^{11} ), \\
  t_{\sigma\tau} &= \textstyle{\frac{1}{16}}
  ( T^{00} -  T^{01} -  T^{10} + T^{11} ), 
\end{align}
\end{subequations}
where we use the notation $T^{ST}$.

The terms
involving the spin cancel out for nucleon scattering off closed-shell targets. For this reason, we pay attention to
$t_0$ and $t_\tau$.
In the case of proton scattering, we need to couple it with
the target protons or neutrons, which we denote by $\tau$.
If $\hat{\rho}_\tau$ denotes the density matrix of species $\tau$, then
the elements we need to incorporate in the construction of the
optical potential are of the type 
$\langle p\tau|\,t\,|p\tau\rangle\,\hat{\rho}_\tau$.
Then, the matrix elements of interest are of the type
$\langle p\tau|t|p\tau\rangle$.
Omitting the spin components of the force, 
we obtain the explicit expressions
\begin{subequations}
\begin{align}
  \langle pp\mid\! t\! \mid pp\rangle 
	\label{T1}
  &= t_0 + t_\tau 
  = \tfrac{1}{4}( T^{01} + 3 T^{11} ), \\
  \langle pn\mid\! t \!\mid pn\rangle 
  &= t_0 - t_\tau
  = \tfrac{1}{8}( T^{00} + T^{01} + 3T^{10} + 3 T^{11} ). 
\end{align}
\end{subequations}
We note that 
$\langle pp\mid\! t\! \mid pp\rangle\!=
\!\langle nn\mid\! t\! \mid nn\rangle$,
only contain isospin states $T=1$, as it is shown in Eq.~\eqref{T1}.

In the calculation of proton-nucleus
 optical potentials we
couple the projectile (proton) with target protons and 
target neutrons,
\begin{equation}
U_{p} =
g_{pp} \otimes \rho_p + g_{pn} \otimes \rho_n. \label{up}
\end{equation}
Making explicit the  $\beta$-dependence in the $g$ matrix,
as in Eq. \eqref{g_rb}, we  get
\begin{align} 
  U_{p} 
  = 
  &(1-\beta) g_{pp}^{(0)}\otimes \rho_p  + 
    \beta g_{pp}^{(1)} \otimes \rho_p  +
             \nonumber \\
  &(1 - \beta) g_{pn}^{(0)} \otimes \rho_n +
    \beta g_{pn}^{(1)} \otimes \rho_n. 
    \label{x2}
\end{align}
The corresponding gradient 
contribution becomes
\begin{align} 
 U^{1}_{p}
= 
       &\left[ (1 - \beta) 
        \frac{\partial g_{pp}^{(0)}}{\partial z}- 
        \frac{\partial \beta}{\partial z}  g_{pp}^{(0)} \right]
        \otimes \rho_p + 
        \nonumber \\
      &\left[ (1-\beta) 
        \frac{\partial g_{pn}^{(0)}}{\partial z} -
        \frac{\partial \beta}{\partial z} g_{pn}^{(0)}  \right] 
        \otimes \rho_n +
        \nonumber \\
       &\left[ \beta
        \frac{\partial g_{pp}^{(1)}}{\partial z} +  
        \frac{\partial \beta}{\partial z}  g_{pp}^{(1)} \right]
        \otimes \rho_p +
        \nonumber \\
       &\left[ \beta 
        \frac{\partial g_{pn}^{(1)}}{\partial z} +  
        \frac{\partial \beta    }{\partial z} g_{pn}^{(1)}\right]
        \otimes \rho_n \,.
        \label{U1p}
\end{align}
In the case of neutron-nucleus optical potentials
the coupling reads
\begin{equation}
U_{n} =  g_{np} \otimes \rho_p + g_{nn} \otimes \rho_n, \label{un}
\end{equation}
with the resulting $U_n^1$ given analogous to that 
in Eq.~\eqref{U1p}.

\section{Applications}

In this section, we present and discuss the results 
for nucleon scattering from $^{208}$Pb and $^{48}$Ca
at low and intermediate energies. 
First, we  address
the accuracy of the calculations 
in the context of the $\delta g$ folding approach to subsequently 
assess the role of isospin asymmetry in the same context. 

To calculate $U_1$ in Eq. \eqref{U1} one needs to evaluate the integral 
\begin{equation}
    z^3\Omega({\bm q},{\bm P};z)= 
    \int d{\bm Q} \,\tilde\rho({\bm Q},{\bm P}) 
    \,\frac{j_1(|\bm Q-\bm q|z)}{|\bm Q-\bm q|z}\;. 
    \label{eq:hfa}
\end{equation}
This is a delicate integral because it involves two functions
that peak at different places. One at $Q=0$, and the other
at $\boldsymbol{q}=\boldsymbol{Q}$. In Appendix E we describe 
the method we developed to calculate this integral reliably. In our computational strategy, we make use of the multipole 
expansion of $\Tilde{\rho}(\boldsymbol{Q},\boldsymbol{P})$ 
of Eq.~\eqref{pmultipole}. 

All \textit{NA} potentials are calculated in momentum space.
For this purpose, the
 infinite nuclear matter
$g$ matrices 
 in the BHF approximation
are calculated at twenty values of $k_F$
uniformly distributed over 0 and $1.6$ fm$^{-1}$ for the SNM, 
and over 0 and $1.9$ fm$^{-1}$ for the PNM.
See Refs.~\cite{Arellano2015,Isaule2016} for further details.
Additionally, shell-model s.p. wavefunctions for the targets 
are obtained from Ref.~\cite{Negele1970}.
The resulting nonlocal $NA$ potentials are then used to
calculate their corresponding scattering observables 
for proton and neutron scattering, making use of the 
SWANLOP package \cite{Arellano2019,Arellano2021}.

\subsection{Folding in the Slater approximation: a test case}

In the context of the $\delta g$ folding, it is demonstrated 
in Refs. \cite{Arellano2007a,Aguayo2008} 
that if the density matrix is represented in the Slater approximation, 
 the optical potential expressed by Eqs.~\eqref{U0} 
and \eqref{U1} is reduced to the ABL folding of
Eq. \eqref{UABLth}. 
This property becomes very useful for cross-checking the 
accuracy of the routines 
we have developed to
incorporate the gradient term in $U_1$. 

As shown in Appendix D, for a given radial density $\rho(z)$ 
the density matrix in the Slater approximation becomes~\cite{Campi1978}
\begin{equation}
        \Tilde{\rho}_{_\textrm{Slater}}(Q;P) = 
        4 \pi \int_0^{\infty} z^2\: dz\:
        j_0 (Qz)\: \rho (z)\: n_z(P),
\end{equation}
where
\begin{equation}
    n_z(P) = \frac{1}{\frac{4}{3}\pi \hat{k}_z^3}\Theta(\hat{k}_z - P).
\end{equation}
The local momentum $\hat{k}_z$ is related to the proton or
neutron local density $\rho(z)$ through
\begin{equation}
    \rho(z) = \frac{\hat{k}_z^3}{3\pi^2}.
\end{equation}
In this limit $U(\boldsymbol{k'},\boldsymbol{k})$
becomes the ABL folding, 
relying only on the radial density $\rho(z)$
We then replace the
$\Tilde{\rho}_{_\textrm{Slater}}(Q;P)$ density matrix 
in the numerical calculation of $U_1$. 
If the numerical implementation is consistent, the resulting potentials should match those from ABL.

\begin{figure}[h]
\begin{center}
  \begin{tikzpicture}
\node[inner sep=0pt] (sun) at (0,0)
    {\includegraphics[width=.35\textwidth,angle=270]{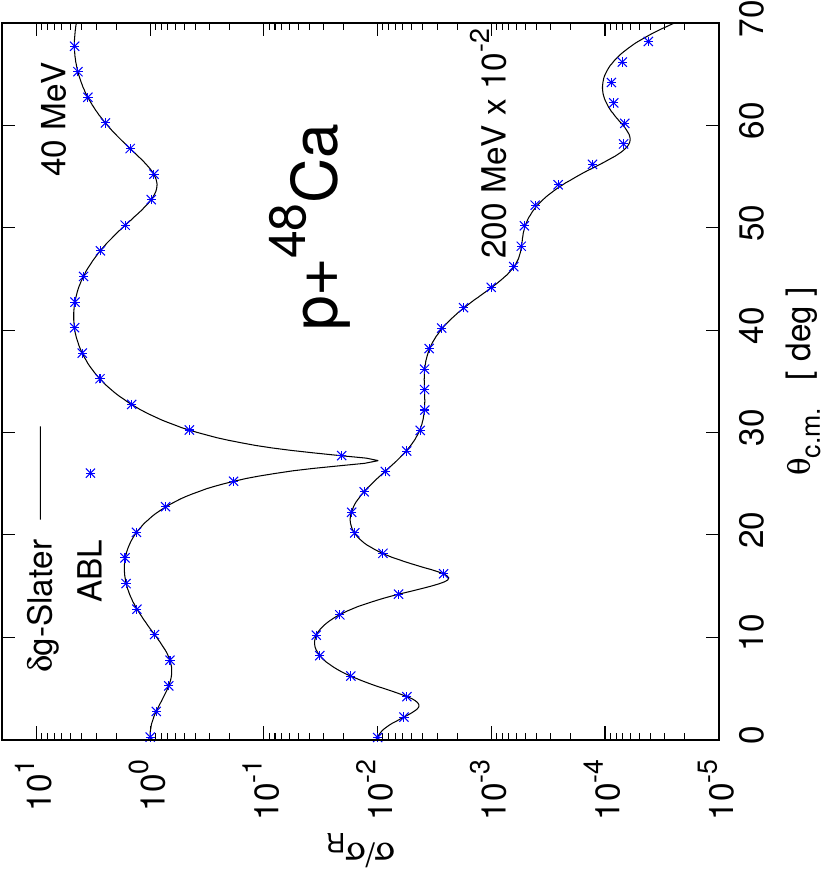}};
\node[inner sep=0pt] (earth) at (6.4,0)
    {\includegraphics[width=.35\textwidth,angle=270]{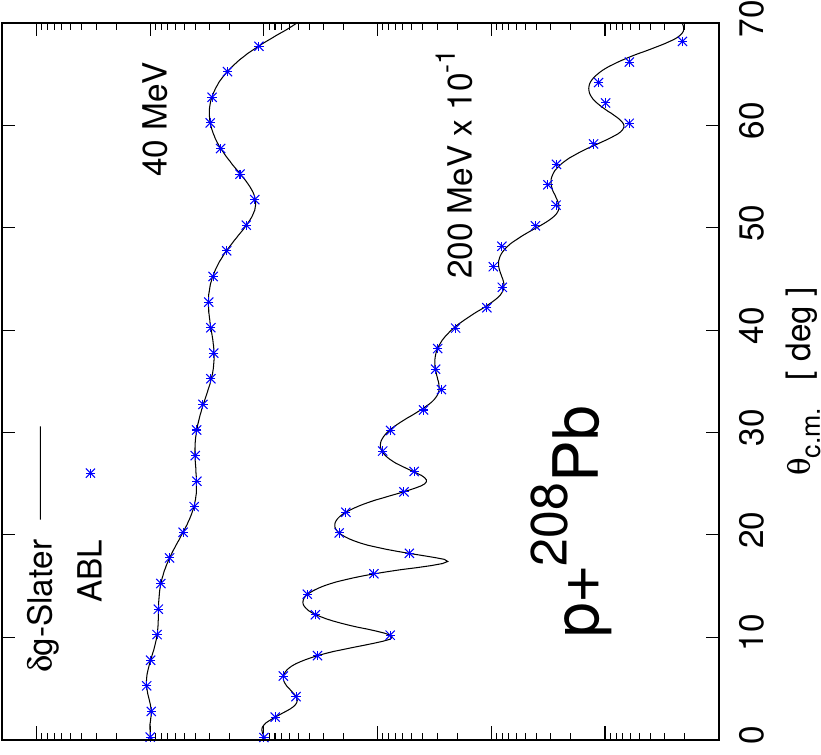}};
\end{tikzpicture}
\end{center} 
    \caption{
    Ratio-to-Rutherford differential cross 
    section for proton elastic scattering from $^{48}$Ca (lhs)
    and $^{208}$Pb (rhs) at 40 MeV and 200 MeV,
    as functions of the c.m. scattering angle.
    Results at 200 MeV for $^{48}$Ca and $^{208}$Pb are
    multiplied by factors of $10^{-2}$ and $10^{-1}$, respectively}.
    \label{fig:ablvsdgs}
\end{figure}

In Fig. \ref{fig:ablvsdgs} we compare differential cross sections 
for optical potentials using the ABL folding (crosses) and
the $\delta g$ folding (solid curves) when the density matrix 
is represented in the Slater approximation. In this figure 
we consider $^{48}$Ca and $^{208}$Pb targets for proton 
at $40$ and $200$~MeV. We observe that the two results 
are identical to the eye.

In Table \ref{tab:table2} we show the volume integral 
per nucleon ($J/A$) for $p+^{48}$Ca, and for $p+^{208}$Pb 
scattering at $40$~MeV. 
Here, we compare ABL and $\delta g$-folding results when the Slater 
approximation is used. 
We note that the differences between the ABL and the $\delta g$ 
calculations are bound by $0.5\%$ at the most, 
a reasonable discrepancy considering the many stages involved 
in the calculations. 
\begin{table}[h]
\caption{Volume integral per nucleon for $p+^{48}$Ca, 
  and $p+^{208}$Pb scattering at 40~MeV.}
\begin{tabular}{ c|cc|cc|}
  \cline{2-5} &\multicolumn{4}{|c|}{ } \\ [-3mm]
              &\multicolumn{4}{|c|}{$J/A$ [ MeV~fm$^3$ ]} \\
              &\multicolumn{4}{|c|}{ } \\ [-3mm]
  \cline{2-5}
              &\multicolumn{4}{|c|}{ } \\ [-3mm]
  &\multicolumn{2}{|c|}{$^{48}$Ca} &\multicolumn{2}{ c|}{$^{208}$Pb} \\
  \cline{2-5}
  &\hspace{4mm}Re\hspace{4mm} &\hspace{4mm} Im\hspace{4mm} &
   \hspace{4mm}Re\hspace{4mm} &\hspace{4mm} Im\hspace{4mm} \\
  \hline
ABL[$g^{(0)}$]         &  -704.5 & -264.3  &  -701.4 &  -237.0 \\
$\delta g$[$g^{(0)}$]  &  -707.2 & -265.3  &  -702.5 &  -237.1 \\ 
                       &          &          &          &       \\[-3mm]
  Difference           &    0.4\% &  0.4\%   &   0.2\%  & 0.1\% \\
  \hline
  \hline
ABL[$g^{(0)}\!+\!g^{(1)}$]  & -633.8 & -220.9  & -611.6 & -188.2  \\
$\delta g$[$g^{(0)}\!+\!g^{(1)}$] & -636.3 & -220.7 & -613.1 & -187.4\\
                       &          &          &          &       \\[-3mm]
Difference                        &  0.4\% & 0.1\% & 0.2\% & 0.5\% \\
  \hline
\end{tabular}
\label{tab:table2}
\end{table}

%
%
\subsection{$\delta g$ folding with isospin-asymmetry}

In this section, we focus on the role of the local isospin asymmetry
$\beta(z)$ in nucleon elastic scattering considering
two different approaches, with two variants each.
The most complete approach is given by the
$\delta g$ folding, where the nonlocal density
    matrix $\Tilde{\rho} (\boldsymbol{Q};\boldsymbol{P})$ is 
    constructed explicitly from shell-model s.p. wavefunctions.
    The second approach is the ABL folding,
    an approximate scheme where the Slater approximation for the 
    density matrix $\Tilde{\rho}(\boldsymbol{Q};\boldsymbol{P})$
    is used. As a result, only radial densities
    $\rho_{p,n}(z)$ are needed.
    For each of these approaches, we consider $\beta$-dependent
    nuclear matter (denoted $g^{(0)}\!+\!g^{(1)}$) and
    SNM (denoted $g^{(0)}$).

\begin{center}
\begin{figure}[ht]
    \includegraphics[width=1\linewidth,angle=0]
    {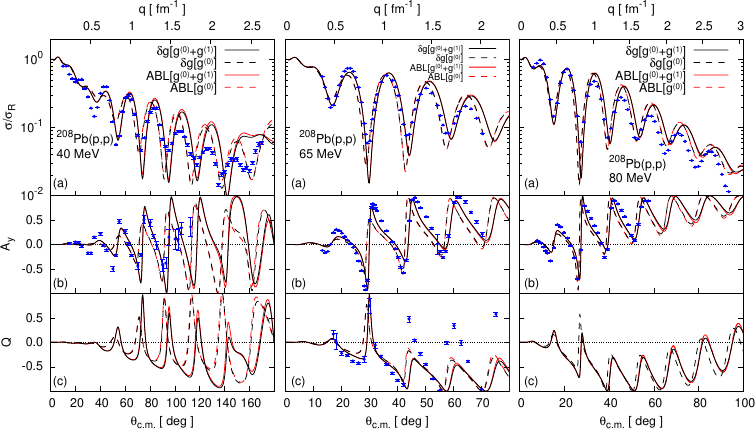}
\caption{ 
    \label{fig:Pb208-All}
    Ratio-to-Rutherford differential cross section
    $\sigma/\sigma_R$ (a), analyzing power $A_y$ (b) and 
    spin rotation function $Q$ (c) for $p+^{208}$Pb scattering
    at 40, 65, and 80 MeV, as a function of the c.m. scattering angle. 
    The upper axis denotes momentum transfer $q$ in units of fm$^{-1}$.
    The data are from Refs. \cite{DataPb208-40MeV,DataPb208-65Mev,DataPb208-80MeV,AyDataPb208-40MeV,AyDataPb208-65MeV,AyDataPb208-80MeV}.
    Solid curves represent results from $\delta g$ folding
    including both $g^{(0)}\!+\!g^{(1)}$. Dashed curves
    represent results considering only $g^{(0)}$ (SNM).
  }
\end{figure}
\end{center}

In Fig. \ref{fig:Pb208-All} we plot the measured and calculated
differential cross section relative to Rutherford cross section
$\sigma/\sigma_R$ (a),
analyzing power $A_y$ (b) and spin rotation function $Q$ (c) for
proton scattering from $^{208}$Pb  at $40$, $65$ and $80$~MeV beam energy.
The data are from 
Refs.~\cite{DataPb208-40MeV,DataPb208-65Mev,DataPb208-80MeV,AyDataPb208-40MeV,AyDataPb208-65MeV,AyDataPb208-80MeV}.
Black curves denote results based on $\delta g$-folding potentials
whereas red curves represent those based on ABL approach.
Additionally, solid curves denote calculations 
treating explicitly isospin asymmetry through $\beta(z)$,
while dashed curves denote results using SNM for the $g$ matrix.

To assess the role of isospin asymmetry at the level of the
\textit{NN} effective interaction,
we pay attention to the low momentum results for the
angular scattering observables, 
namely those with momentum transfer up to about 1.5~fm$^{-1}$.
This criteria is motivated by the $t\rho$ structure of the 
optical potential,
giving hints on the role of the two main inputs
participating the construction of the potential: the target
form factor and the effective interaction.
Hence, if we assume better control on the effective 
interaction and the target densities at low $q$, then the low-$q$ 
scattering observables would become more reliable. 
On the other end, the high-momentum behavior of the density
$\tilde\rho(q)$ is more uncertain, therefore any disagreement
between the model and the data can be attained to such uncertainties, 
in addition to higher order effects.
%

Results at 40~MeV in Fig. \ref{fig:Pb208-All} exhibit clear differences
between calculations including $g^{(1)}$ (solid curves) 
and excluding it (dashed curves) in the $g$ matrix,
feature observed at all angles.
Additionally, the differences between $\delta g$- and ABL-folding 
approaches are relatively small.
When comparing to the data we observe that the inclusion of $g^{(1)}$ 
favors an agreement with the data, feature more noticeable at
momentum transfers $q\lesssim 1$~fm$^{-1}$,
for $\sigma/\sigma_R$ and analyzing power $A_y$.
At 65~MeV and 80 MeV we note the same tendency, 
although the role of $g^{(1)}$ becomes weaker
as the energy increases.

Another feature also noted in Fig.~\ref{fig:Pb208-All} is the
similarity between $\delta g$ and ABL folding when the
same considerations are applied to the $g$ matrix. 
This indicates that the optical potential is not quite sensitive
to the representation of the density matrix.
This statement is valid as long as we use the same starting
energy $\omega$ in the $g$ matrix for each approach. 
As stated earlier, in this work we use
as starting energy $\omega\!=\!E + \bar e$,
with $\bar e$ the nuclear-matter average s.p. energy 
at the local density $\rho(z)$.
However, in a Hartree-Fock shell-model for the target each 
shell $\alpha$ has its corresponding s.p. energy $\varepsilon_\alpha$. 
Hence, we should perform folding calculations for each shell
in the nucleus, each one with its associated effective interaction
at $\omega_\alpha\!=\!E\!+\!\varepsilon_\alpha$.
Although this is an interesting issue to investigate,
its implementation and study is computationally more demanding,
beyond the scope of the present work.
%
%

\begin{center}
  \begin{figure}
    \includegraphics[width=0.8\linewidth,angle=-90]
    {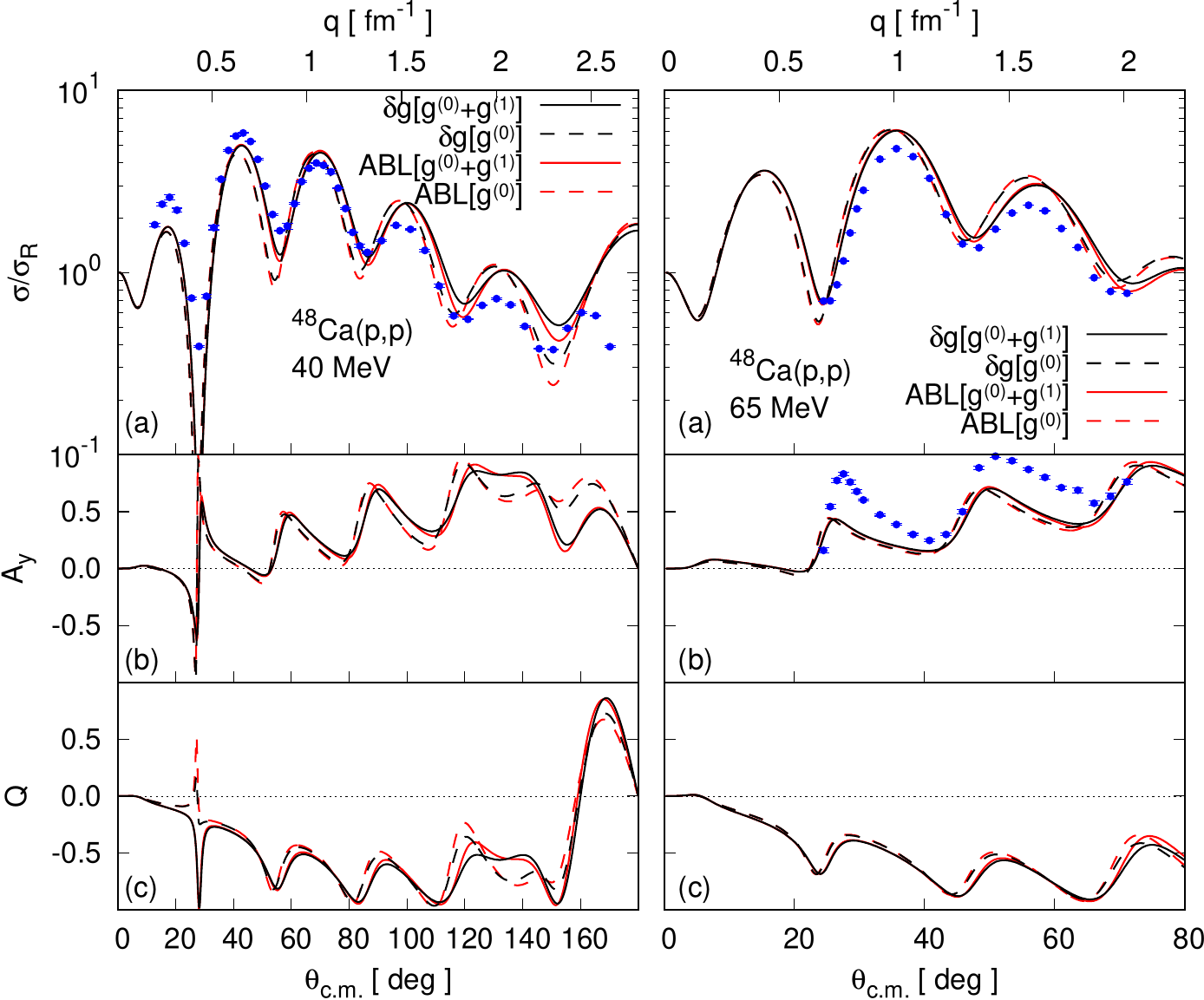}
\caption{ 
  \label{fig:ca48-40/65}
Ratio-to-Rutherford cross section $\sigma/\sigma_R$ (a),
analyzing power $A_y$ (b) and spin rotation function $Q$ (c) for
$p\!+\!^{48}$Ca elastic scattering at $40$ and $65$~MeV.
  The data are from Refs. \cite{DataCa48-40MeV,DataCa48-65MeV}.
    Curve textures follow the same convention as in 
    Fig.~\ref{fig:Pb208-All}.
  }
  \end{figure}
\end{center}

In Fig. \ref{fig:ca48-40/65} we 
plot the measured~\cite{DataCa48-40MeV,DataCa48-65MeV}
and calculated
ratio-to-Rutherford cross section 
$\sigma/\sigma_R$ (a),
analyzing power $A_y$ (b) and spin rotation function $Q$ (c) for
proton elastic scattering from $^{48}$Ca 
 at $40$ and $65$~MeV beam energies.
We follow the same convention for the curves
as in Fig. \ref{fig:Pb208-All}.
Here we also evidence better agreement between the 
data at 40 MeV and both folding approaches 
including $g^{(0)}$ and $g^{(1)}$, 
at momentum transfer $q \lesssim 1$~fm$^{-1}$.
In the case of 65~MeV, however, differences between
the two approaches are less pronounced, 
implying weaker effects stemming from the isospin asymmetry 
at the level of the \textit{NN} effective interaction.
%

\begin{center}
  \begin{figure}
    \includegraphics[width=0.8\linewidth,angle=-90]
    {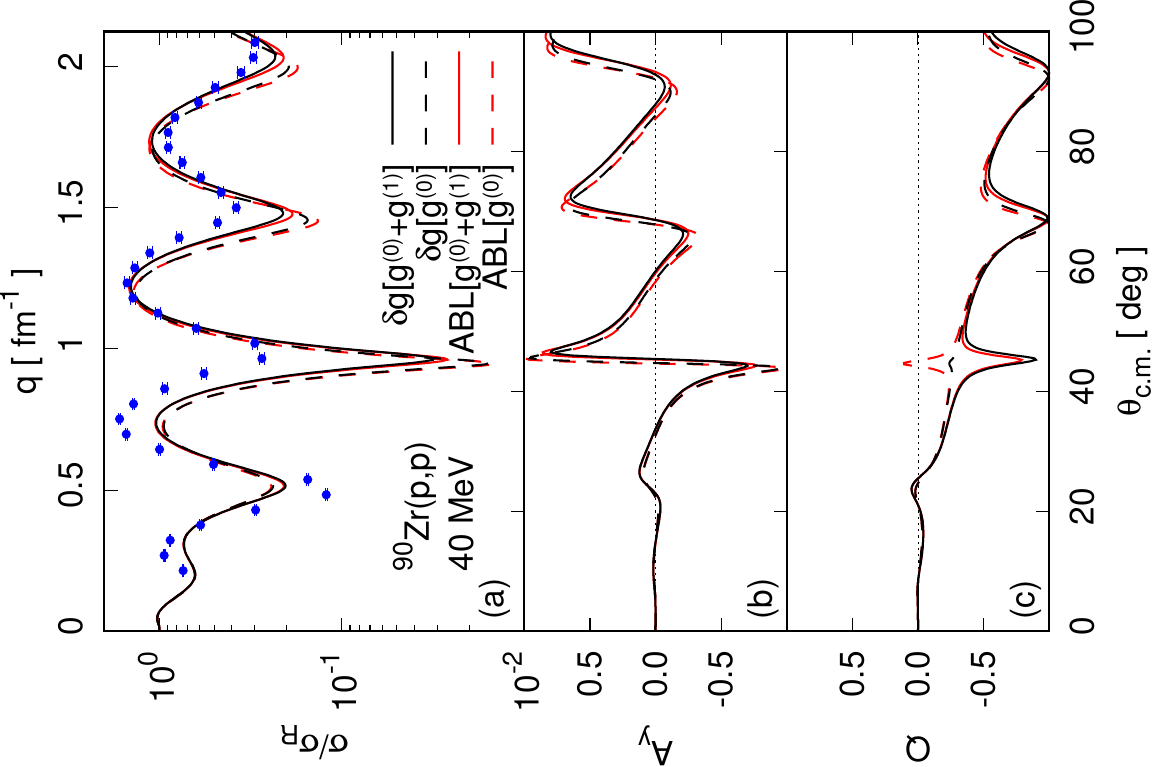}
\caption{ 
  \label{fig:zr90-40}
Ratio-to-Rutherford cross section $\sigma/\sigma_R$ (a),
analyzing power $A_y$ (b) and spin rotation function $Q$ (c) for
$p\!+\!^{90}$Zr elastic scattering at $40$~MeV.
    The data are from Refs. \cite{DataZr90-40MeV}.
    Curve textures follow the same convention as in Fig.~\ref{fig:Pb208-All}.
  }
  \end{figure}
\end{center}

 We have also investigated $p\!+\!^{90}$Zr elastic scattering 
 including and suppressing asymmetric nuclear matter contributions.
 In this case, the isospin asymmetry $\beta(z)$ for the target is rather small,
 as observed in Fig.~\ref{fig:RhoDBeta}. 
 Note that in this case $\beta(z)$ is 0.2 at the most.
 Additionally, $\beta(z)$ is the smallest at the interior of the nucleus
 ($z\!<\!z_{rms}$), when compared with $^{208}$Pb and $^{48}$Ca.
 In Fig.~\ref{fig:zr90-40} we show results for $\sigma/\sigma_R$, $A_y$ and $Q$  
 at beam energy of 40~MeV, an energy where we expect the effect
 of $g^{(1)}$ gets most pronounced. 
 The curve textures follow the same convention as in Fig.~\ref{fig:Pb208-All}.
 As observed, the differences between solid and dashed curves is
 quite moderate, with a slight shift of $\sigma/\sigma_R$ 
 to forward angles of the first diffractive minimum.
%
%

\begin{center}
  \begin{figure}[ht]
    \includegraphics[width=0.8\linewidth,angle=-90]
    {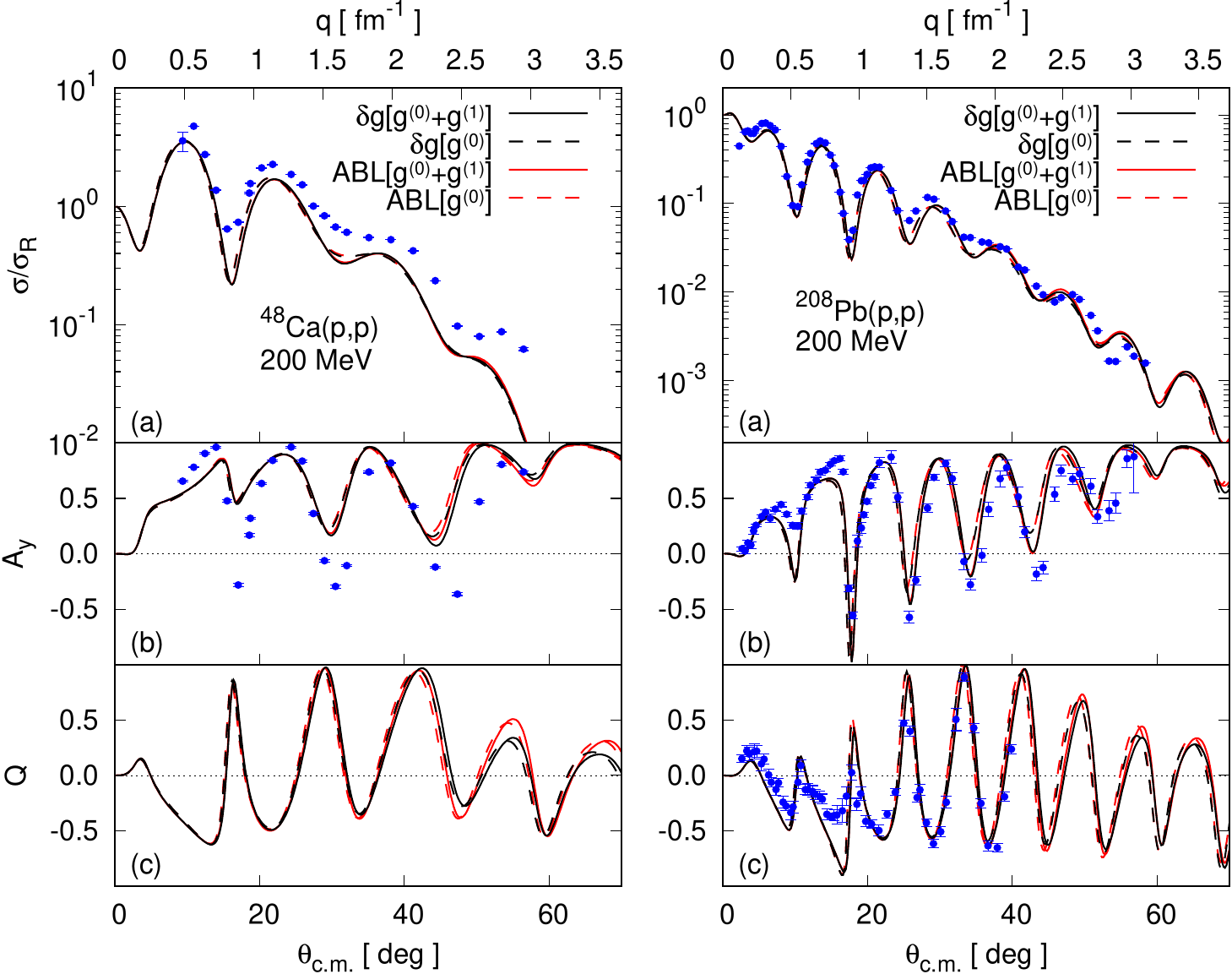 }
\caption{ 
  \label{fig:xaq_200}
  Ratio-to-Rutherford differential cross section 
  for $p+^{208}$Pb and $p+^{48}$Ca scattering at 200 MeV 
  beam energy, as a function of the c.m. scattering angle. 
The data are from Refs. \cite{DataCa48_200MeV,DataPb208-200MeV}.
  }
  \end{figure}
\end{center}

We have extended this study to higher energies. 
In Fig.~\ref{fig:xaq_200} we plot the 
ratio-to-Rutherford differential cross section
$\sigma/\sigma_R$ (a),
analyzing power $A_y$ (b) and spin rotation $Q$ (c),
as functions of the scattering angle for proton elastic 
scattering from $^{208}$Pb and $^{48}$Ca.
These results show that the inclusion of $g^{(1)}$ in the
\textit{NN} effective interaction results in
marginal effects in the resulting scattering observables.
This result is consistent with findings reported
in Ref.~\cite{Chinn:1993zza}
by Chinn {\it et al.}, 
where the Watson implementation of the multiple scattering 
expansion was contrasted with the KMT implementation, 
resulting in no visible differences at beam
energies of 200~MeV and above.

\begin{center}
  \begin{figure}[ht]
    \includegraphics[width=0.9\linewidth,angle=-90]
    {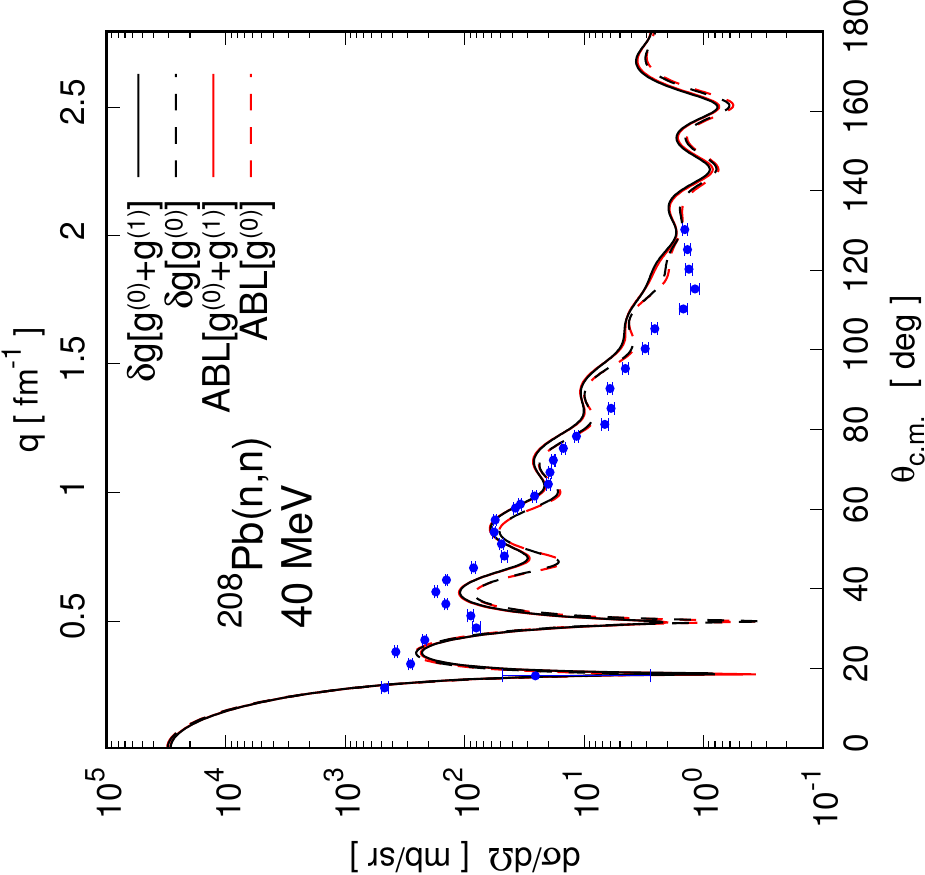 }
\caption{ 
  \label{fig:nPb208-40}
  Differential cross section for $n+^{208}$Pb 
scattering at 40 MeV beam energy as a function of 
    the c.m. scattering angle. 
The data are from Ref. \cite{DataPb208n-40MeV}.
  }
  \end{figure}
\end{center}

\begin{center}
  \begin{figure}[ht]
    \includegraphics[width=0.9\linewidth,angle=-90]
    { 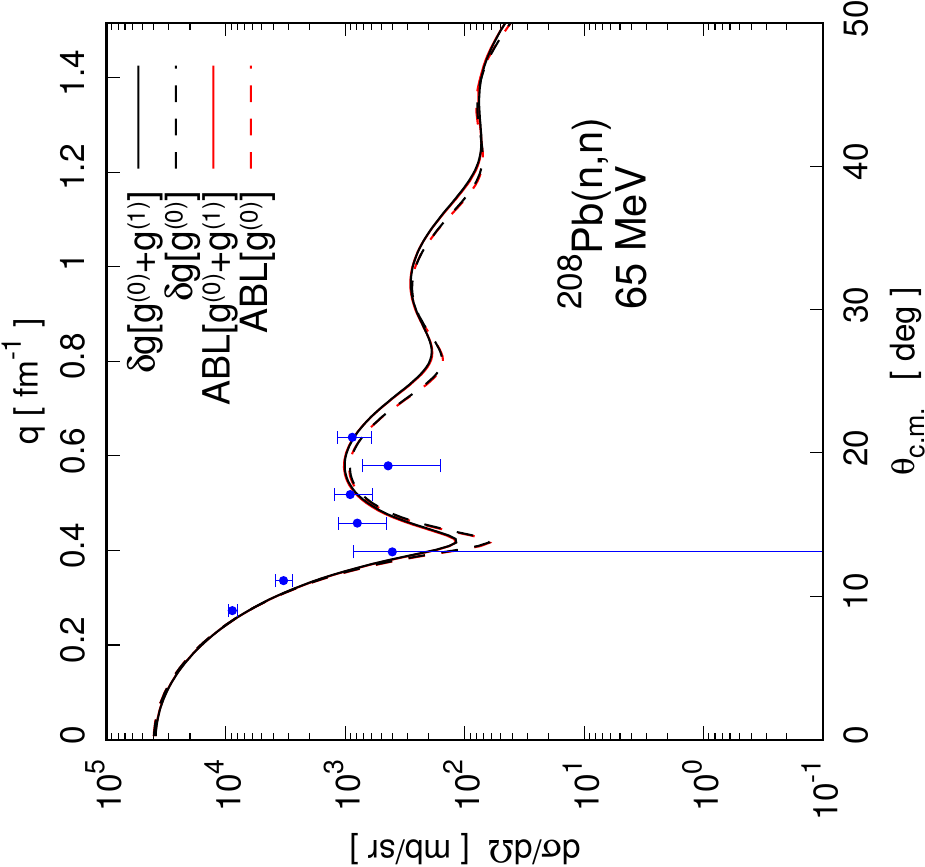 }
\caption{ 
  \label{fig:nPb208-65}
  Differential cross section for $n+^{208}$Pb 
scattering at 65 MeV as a function of the c.m. scattering angle. 
The data are from Ref. \cite{DataPb208n-65MeV}.
  }
  \end{figure}
\end{center}

\begin{center}
  \begin{figure}[ht]
    \includegraphics[width=0.9\linewidth,angle=-90]
    { 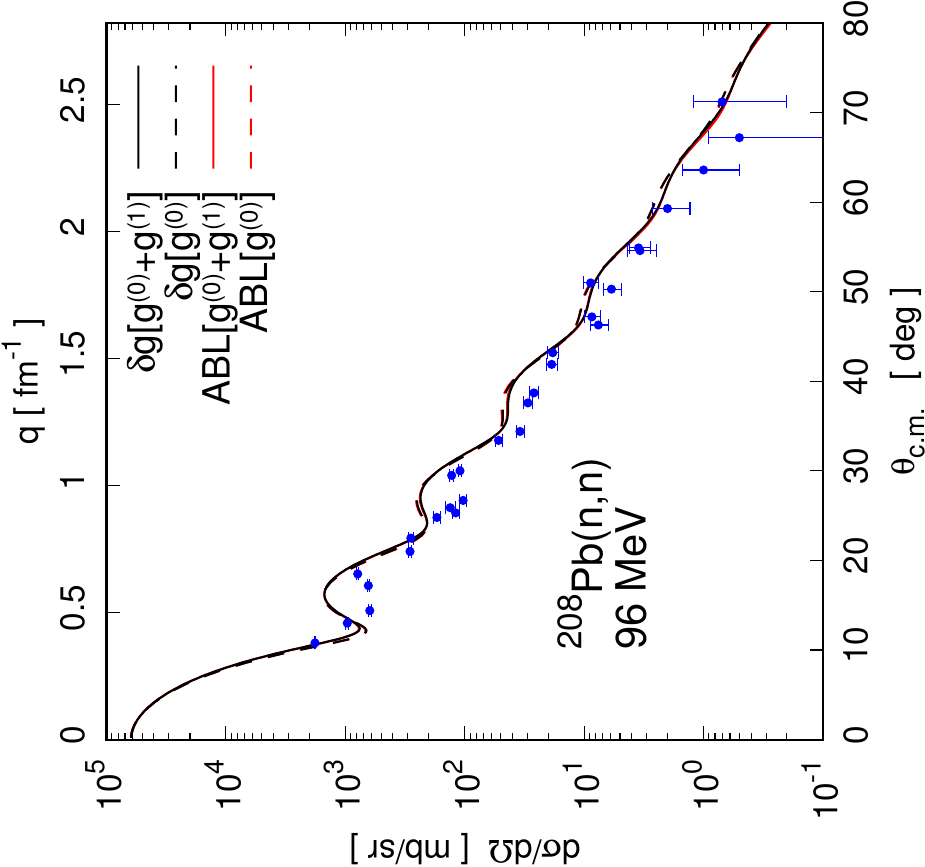 }
\caption{ 
  \label{fig:nPb208-96}
  Differential cross section for $n+^{208}$Pb scattering 
  at 96 MeV as a function of the c.m. scattering angle. 
  The data are from Ref. \cite{DataPb208n-96MeVb}.
  }
  \end{figure}
\end{center}

We now investigate the role of $g^{(1)}$ in the case of 
neutron scattering. 
In Figs. \ref{fig:nPb208-40}, \ref{fig:nPb208-65}, 
and \ref{fig:nPb208-96} we show differential cross sections
as functions of the scattering angle for $n\!+\!^{208}$Pb
elastic scattering at $40$, $65$ and $96$~MeV, respectively. 
 We follow the same convention for black and red curves 
 as in Fig. \ref{fig:Pb208-All}. 
 As in the cases of proton scattering,
 we observe small differences between 
 $\delta g$ and ABL folding approaches.
 Additionally, we also note that scattering at 40~MeV 
 is more sensitive to the inclusion of the isovector 
 part of the $g$ matrix,
 feature which almost disappears at 96~MeV.
 The agreement with the data, however, is quite modest.
 At 40~MeV the calculated $d\sigma/d\Omega$ is much
 more diffractive than the data,
 at 65~MeV the calculated $d\sigma/d\Omega$ underestimates
 the data, whereas at 96~MeV the calculated $d\sigma/d\Omega$
 appears more diffractive than the data.
 Whether these shortcomings have a common ground
 is an issue which remains to be investigated.
%
%
%

Total cross sections are among the properties of 
interest in \textit{NA} scattering.
In Table \ref{tab:table3} we show the calculated elastic ($\sigma_e$), 
reaction ($\sigma_R$) and total ($\sigma_T$) cross section 
for neutron elastic scattering from $^{48}$Ca 
and $^{208}$Pb at $40$~MeV. Here, we include all four 
approaches discussed in the preceding discussion. 
We observe that the overall discrepancy among all the 
calculated cross section is within $2-3\%$. 
In light of these results, we can state that ABL 
folding with SNM is a reasonable approach for 
calculating total cross sections, 
at least at energies at and above $40$~MeV.  
The experimental value for the total cross section of 
$n+^{208}$Pb at 40~MeV is 4.38~b \cite{nPb208txs}. 
The inclusion of $g^{(1)}$ in both 
approaches gets us closer to this value.

\begin{table}[h]
\caption{Calculated elastic, reaction and total cross section for 
  $n+^{48}$Ca and $n+^{208}$Pb scattering at $40$~MeV.}
\begin{tabular}{c|cc|cc|}

%
%
  \cline{2-5}
              &\multicolumn{4}{c|}{ } \\ [-4mm]
  &\multicolumn{2}{c|}{ABL} &\multicolumn{2}{ c|}{$\delta g$} \\
 \cline{1-5} 
  \multicolumn{0}{|c|}{$n+^{48}$Ca} &\hspace{1mm}$g^{(0)}$\hspace{1mm} &\hspace{1mm}   $g^{(0)}+g^{(1)}$\hspace{1mm} &
   \hspace{1mm}$g^{(0)}$\hspace{1mm} &\hspace{1mm} $g^{(0)}+g^{(1)}$ \hspace{1mm} \\
   \hline 
\multicolumn{1}{|c|}{$\sigma_T$ [ b ]  } 
  &    $2.19$     &    $2.17$       &  $2.17$     &   $2.15$\\
\multicolumn{1}{|c|}{$\sigma_e$ [ b ] }   
  &    $1.13$     &    $1.17$       &  $1.12$     &   $1.16$       \\
\multicolumn{1}{|c|}{$\sigma_R$ [ b ]  }   
  &    $1.06$     &    $1.00$       &  $1.05$     &   $0.994$       \\
\hline \hline 
  \multicolumn{0}{|c|}{$n+^{208}$Pb} &\hspace{1mm}\hspace{1mm} &\hspace{1mm}   \hspace{1mm} &
   \hspace{1mm}\hspace{1mm} &\hspace{1mm}  \hspace{1mm} \\
   \hline 
\multicolumn{1}{|c|}{$\sigma_T$ [ b ]}    
  &    $4.99$    &    $4.84$       &  $4.96$     &   $4.82$ \\
\multicolumn{1}{|c|}{$\sigma_e$ [ b ]}    
  &    $2.39$    &    $2.33$       &  $2.37$     &   $2.32$       \\
\multicolumn{1}{|c|}{$\sigma_R$ [ b ]}   
  &    $2.60$    &    $2.51$       &  $2.59$     &   $2.50$       \\

\hline
\end{tabular}
\label{tab:table3}
\end{table}

\section{Summary and Conclusions}

We have investigated the role of isospin asymmetry
in the \textit{NN} effective interaction
in the context of \textit{NA} elastic scattering.
To this purpose we assume that the \textit{in-medium} $g[\rho,\beta]$ 
matrix is an admixture of isospin-symmetric nuclear matter
and pure neutron matter solutions of BHF equations, 
with a linear dependence on the asymmetry parameter $\beta$.
This construction is supported by studies 
on asymmetric nuclear matter reported by Bombaci 
\textit{et al.}~\cite{Bombaci1991}.
We base this study on the Argonne $v_{18}$ bare potential 
due to its ability to describe \textit{NN} scattering  up to 350~MeV.
The isospin-asymmetric $g$ matrices are then used for
the construction of
optical model potentials for elastic nucleon scattering 
off closed-shell nuclei.
For this purpose, we make use of the $\delta g$ folding approach 
introduced in Ref.~\cite{Arellano2007a},
suited for an explicit treatment of the nonlocal density matrix.
In this work, we make use of shell-model s.p. wavefunctions
in a Hartree-Fock description of the ground state of nuclei.
The folding approach we pursue 
provides us with a well-defined prescription to track the
local isospin asymmetry throughout the target
utilizing the
\textit{in-medium} $g[\rho,\beta]$ matrix.
The calculated optical potentials are nonlocal, 
as all intrinsic nonlocalities in the $g$ matrix are retained. 
Elastic scattering observables are obtained by employing
 the {\small SWANLOP} package~\cite{Arellano2021},
suited for nonlocal potentials in momentum
space, including the Coulomb interaction.

We have found that the inclusion of isospin asymmetry in the $g$ matrix, 
characterized through the local isospin asymmetry $\beta(z)$ and the 
inclusion of its isovector part, yields a reasonable description of 
differential cross sections at nucleon beam energies 
between 40 and 100~MeV.
We also find that the use of the fully nonlocal density matrix,
within the $\delta g$ folding approach, yields
similar results to those obtained within the ABL approach.
This finding is useful in the sense that the calculation of optical
potentials in the ABL approach is computationally less intensive
than the $\delta g$ folding approach.
Thus, the ABL folding model constitute a safe starting approach to
investigate \textit{NA} collisions.

For proton scattering at energies of 65~MeV and below,
the inclusion of $g^{(1)}$ yields better agreement with the scattering
data at momentum transfers up to $\sim$1~fm$^{-1}$,
relative to approaches based on SNM $g$ matrices.
The physical implication of this result is that the
isospin-asymmetry in the \textit{NN} effective interaction
is not negligible for a complete description of 
\textit{NA} scattering at these energies.
As the energy increases, the role of $g^{(1)}$ becomes weaker. 

We conclude that the treatment of the local isospin asymmetry 
in the nucleus 
and associated site-dependent $g[\rho,\beta]$ matrix play an 
important role in 
the physics of collision processes involving isospin asymmetric targets. 
To assess the limits of these findings, it would be interesting to
consider alternative bare \textit{NN} interactions such as 
chiral potentials, and
other solutions in the representation of the target ground state, 
in addition to open-shell nuclei as targets.
An additional issue of interest is to investigate how the 
nonlocality of the optical potential gets expressed in its 
isovector component.
In Ref.~\cite{Arellano2024a}, the authors
introduce the so-called $JvH$ universal factorization of the 
optical potential.
The question would then be to characterize the nonlocality of
the isovector term.
These issues constitute natural extensions of this work.

\begin{acknowledgments}
We are grateful to Prof. Ch. Elster for her valuable help 
in the preparation of this manuscript.
The authors thank the referee for her/his
thorough revision of the manuscript as well as asserted
remarks and observations. 
H.F.A. acknowledges partial support provided by the supercomputing 
infrastructure of the NLHPC (ECM-02): Powered@NLHPC.
He also thanks the hospitality of colleagues of CEA-DAM at
Bruy\`eres-le-Ch\^atel, France, where part of this work was done.
\end{acknowledgments}

\appendix
\section{Slater approximation}
	
The density matrix in a momentum representation
$\tilde\rho({\bm p}',{\bm p})$
can be related
to its  coordinate representation
$\rho({\bm r'},{\bm r})$ by a Fourier transform.
In terms of the momentum variables defined by Eqs.~\eqref{PQ} 
it can be shown 
that $\tilde\rho$ can be
expressed as
\begin{align}
    \tilde\rho(
    {\bm P} + \textstyle{\frac{1}{2}}\,{\bm Q}\,,
    {\bm P} - \textstyle{\frac{1}{2}}\,{\bm Q}) =& 
    \frac{1}{(2 \pi)^3} \int d{\bm z}\, d{\bm s}\,
    e^{i{\bm z} \cdot {\bm Q}}\,
    e^{i{\bm s} \cdot {\bm P}}\,\times
    \nonumber \\
    &
    \rho(
    {\bm z} + \textstyle{\frac{1}{2}}\,{\bm s}\,,
    {\bm z} - \textstyle{\frac{1}{2}}\,{\bm s})\:. 
    \label{rhomomentum}
\end{align}
In this expression ${\bm z}$ refers to
the mean struck-nucleon coordinate $({\bm r}\!+\!{\bm r'})/2$, 
quantity denoted as $\bm R$ in Ref.~\cite{Arellano1990b}.
Additionally, ${\bm s}$ represents the difference between
${\bm r}'$ and ${\bm r}$.

Campi and Bouyssy~\cite{Campi1978} have shown that to a very good
approximation, the 
density matrix in coordinate space for  protons 
or neutrons can be cast in the Slater approximation as
\begin{equation}
    \rho(
    \boldsymbol{z} + \frac{1}{2}\boldsymbol{s}\,,
    \boldsymbol{z} - \frac{1}{2}\boldsymbol{s}) 
    \approx 
    \rho(z) \,F(z;s). \label{rho_slater}
\end{equation}
In this approximation, $\rho(z)$ 
represents the point density 
given by the diagonal terms of the
coordinate-space density matrix,  with $F(z;s)$ being
a measure of the non-locality of the density matrix. 
To ensure that the matter form factor is correctly reproduced, 
the condition $F(z;0)=1$ must be satisfied.
A feature implied by this approximation
is the independence of the density matrix on the angle 
between $\boldsymbol{z}$ and $\boldsymbol{s}$, resulting in an
independence of $\tilde\rho$ on $\hat{\bm Q}\!\cdot\!\hat{\bm q}$ 
in Eq.~\eqref{rhomomentum}.

Following Campi and Bouyssy, 
for $F(z;s)$ the structure suggested by nuclear matter,
namely
\begin{equation}
    F(z;s) = 3\,\frac{j_1(\hat{k}(z)\,s)}{\hat{k}(z)\,s}, \label{slatter}
\end{equation}
is assumed,
where $j_1$ is the spherical Bessel function of order 1. 
The choice 
for $\hat{k}(z)$ depends on the level of approximation 
required for the density matrix. The simplest choice corresponds 
to the Slater approximation
\begin{equation}
   \hat{k}(z) \to
   \hat{k}_{_\textrm{Slater}}(z) = [\,3 \pi ^2\: \rho(z)\,]^{1/3}.
   \label{Slater}
\end{equation}
Thus, the momentum-space density matrix becomes
\begin{equation}
    \tilde\rho(
    {\bm P} + \textstyle{\frac{1}{2}}{\bm Q},
    {\bm P} - \textstyle{\frac{1}{2}}{\bm Q}) 
    \approx 
    \displaystyle{\frac{1}{(2 \pi)^3}} \int d{\bm z}\: 
 e^{i{\bm z} \cdot {\bm Q}} \:\rho(z)\:G(z;P)\:,
\end{equation}
where 
\begin{equation}
    G(z;P) = \int d\boldsymbol{s}\: e^{i{\bm s} \cdot {\bm P}}\: F(z;s)\;.
    \label{GRP}
\end{equation}
Using Eq.~\eqref{Slater} for $\hat{k}(z)$ yields
\begin{equation}
    G(z;P) = \frac{2}{\hat{\rho}(z)}\Theta[\hat{k}(z) - P]\;,
\end{equation}
with $\hat{\rho}(z)$ 
given by
\begin{equation}
    \hat{\rho}(z)=\frac{\hat{k}(z)^3}{3\pi^2}. 
    \label{rho_kfermi}
\end{equation}
After substituting Eq.~\eqref{GRP} for 
$G(R;P)$ into Eq.~\eqref{rhomomentum} we obtain
$\tilde\rho\to\tilde\rho_\textrm{Slater}$, with
$\tilde\rho_\textrm{Slater}$ expressed as
\begin{equation}
        \Tilde{\rho}_{_\textrm{Slater}}(Q;P) = 
        4 \pi \int_0^{\infty} z^2\: dz\:
        j_0 (Qz)\: \rho (z)\: n_z(P),
\end{equation}
where we have defined
\begin{equation}
    n_z(P) = 
    \frac{1}{\frac{4}{3}\pi \hat{k}_z^3}\:\Theta(\hat{k}_z - P)\:,
    \label{nzP}
\end{equation}
with $\hat k_z\!\equiv\!\hat k(z)$. Note that in this limit
case the density matrix depends uniquely on the radial density $\rho(z)$.


\section{Evaluation of $z^3 \Omega (\textbf{q},\textbf{P};z)$}
In Eq. \eqref{eq:hfa} we need to evaluate 
the three-dimensional integral
\begin{equation}
\label{z3omega}
z^3\Omega({\bm q},{\bm P};z) =
  \int d{\bm Q} \:
\tilde\rho({\bm Q};{\bm P}) \:
\frac{j_1(|\bm Q-\bm q|z)}{|\bm Q-\bm q|z}\;,
\end{equation}
where the density matrix $\tilde\rho(\bm Q;\bm P)$ depends,
in the general case, on the angle between $\hat{\bm Q}$
and $\hat{\bm P}$.
We present here
a method to evaluate it with accuracy. 
Let us express $\tilde\rho$ as a sum of multipoles of the form
\begin{equation}
\label{multipole}
\tilde\rho({\bm Q};{\bm P}) =
  \sum_{n} 
  \tilde\rho_n(Q,P)\,P_n(w)\;,
\end{equation}
with $P_n(w)$ being the Legendre polynomials and 
$w\!=\!\hat{\bm Q}\cdot\hat{\bm P}$.
In addition, we recall the expression of the Fourier transform
of a three-dimensional hard sphere, where
\begin{equation}
  \label{well}
  \int d{\bm x}\, 
  \Theta(z-x)\,
  e^{i \bm Q\cdot\bm x} =
  4\pi z^3\,\frac{j_1(Qz)}{Qz}\;,
\end{equation}
where $\Theta$ is the usual Heaviside step function.
After replacing Eqs. \eqref{multipole} for $\tilde\rho$ and
Eq.~\eqref{well} for $j_1(t)/t$ into Eq.~\eqref{z3omega} for
$z^3\Omega$ we obtain
\begin{align}
z^3\Omega({\bm q},{\bm P};z) =&
  \frac{1}{4\pi}\, \sum_n
  \int d{\bm Q}
  \int d{\bm x}\:
  \Theta(z-x)\times \nonumber \\
  &e^{i (\bm Q-\bm q)\cdot\bm x} 
  \tilde\rho_n(Q,P)\,P_n(\hat{\bm Q}\cdot\hat{\bm P})\;.
  \label{z3w_2}
\end{align}
Using
spherical coordinates for $\boldsymbol{x}$, 
where $d\boldsymbol{x}\!=\!x^2\:dx\:d\Omega_x $,
the solid-angle integral in $d\Omega_x$ can be performed 
analytically. Indeed, after expanding in partial waves 
$e^{i\bm Q\cdot\bm x}$ and
$e^{-i\bm q\cdot\bm x}$, the solid angle integration yields
\begin{equation}
  \label{dhatx}
  \int d\Omega_x \,
  e^{i \bm Q\cdot\bm x} 
  e^{-i \bm q\cdot\bm x}  =  
  4\pi\sum_{m} (2m+1) j_m(Qx)\,j_m(qx)\,
  P_m(\hat{\bm Q}\cdot\hat{\bm q})\;.
\end{equation}
This is substituted in Eq. \eqref{z3w_2}, leading to
\begin{align}
  \label{z3w_3}
  z^3\Omega({\bm q},{\bm P};z) =&
  \sum_{n}\sum_{m}
  \int d{\bm Q} \,
  \int x^2\,dx\,
  \Theta(z-x)\,\times\nonumber \\
  &(2m+1) j_m(Qx)\,j_m(qx) \times \nonumber \\
  &
  \tilde\rho_n(Q,P)\,
  P_m(\hat{\bm Q}\cdot\hat{\bm q})\,
  P_n(\hat{\bm Q}\cdot\hat{\bm P})\;.
\end{align}
Considering the identity
\begin{equation}
  \label{identity}
  \int d\hat{\bm Q} \,
  P_m(\hat{\bm Q}\cdot\hat{\bm q}) \,
  P_n(\hat{\bm Q}\cdot\hat{\bm P})  =
  \frac{4\pi}{2n+1} \,
  P_n(\hat{\bm q}\cdot\hat{\bm P})
  \,\delta_{nm}\,,
\end{equation}
Eq.~\eqref{z3w_3}  results in
\begin{align}
  \label{z3w_4}
  z^3\Omega({\bm q},{\bm P};z) =&
  4\pi \sum_{n}
  \int Q^2\,dQ\,
  \int x^2\,dx\,
  \Theta(z-x)\,\times\nonumber \\
  &
  j_n(Qx)\,j_n(qx) \times \nonumber 
  \tilde\rho_n(Q,P)\,
  P_n(\hat{\bm q}\cdot\hat{\bm P})\;.
\end{align}
In this way, if we define the hybrid representation 
of the density matrix by
\begin{equation}
  \label{hybrid}
  \bar{\bar{\rho}}_n(x,P) =
  \int Q^2\,dQ\,j_n(Qx)\,\tilde\rho_n(Q,P)\;,
\end{equation}
and Eq.~\eqref{z3omega} for $z^3\Omega$ 
reduces to
\begin{equation}
  \label{z3w_5}
  z^3\Omega({\bm q},{\bm P};z) =
  4\pi \sum_{n}
  \int_0^z x^2\,dx\,
  j_n(qx) 
  \bar{\bar\rho}_n(x,P)\,
  P_n(\hat{\bm q}\cdot\hat{\bm P})\;.
\end{equation}
The advantage of this expression over that in 
Eq.~\eqref{z3omega} is that it requires to perform only two well
controllable integrals, avoiding the direct angular integration
over the solid angle $d\Omega_Q$.
This expression has been implemented in the 
computational codes to evaluate the gradient term of $U_1$.


%
 \end{document}